\documentclass[journal=ancac3,manuscript=article]{achemso}

\usepackage{xr}
\usepackage{subcaption}
\usepackage{amsmath} 
\usepackage{amssymb}
\usepackage{gensymb}
\usepackage{multirow}
\usepackage{xcolor}
\usepackage{siunitx}

\makeatletter
\newcommand*{\addFileDependency}[1]{
  \typeout{(#1)}
  \@addtofilelist{#1}
  \IfFileExists{#1}{}{\typeout{No file #1.}}
}
\makeatother

\newcommand*{\myexternaldocument}[1]{
    \externaldocument{#1}
    \addFileDependency{#1.tex}
    \addFileDependency{#1.aux}
}

\myexternaldocument{SI}

\author{Alexander V. Petrunin}
\affiliation{Institute of Experimental Colloidal Physics, Heinrich-Heine University, Universit\"{a}tsstr. 1, D\"{u}sseldorf, Germany}
\alsoaffiliation[IPC]
{Institute of Physical Chemistry, RWTH Aachen University, Landoltweg 2, 52074 Aachen, Germany, EU}
\altaffiliation{These authors contributed equally to this work.}

\author{Susanne Braun}
\affiliation[DWI]
{DWI – Leibniz Institute for Interactive Materials, Forckenbeckstr. 50, 52074 Aachen, Germany, EU}
\alsoaffiliation[ITMC]
{Functional and Interactive Polymers, Institute of Technical and Macromolecular Chemistry (ITMC), RWTH Aachen University, Worringerweg 2, 52074 Aachen, Germany, EU}
\altaffiliation{These authors contributed equally to this work.}

\author{Felix J. Byn}
\affiliation[IPC]
{Institute of Physical Chemistry, RWTH Aachen University, Landoltweg 2, 52074 Aachen, Germany, EU}

\author{Indr\'{e} Milvydait\'{e}}
\affiliation[DWI]
{DWI – Leibniz Institute for Interactive Materials, Forckenbeckstr. 50, 52074 Aachen, Germany, EU}
\alsoaffiliation[ITMC]
{Functional and Interactive Polymers, Institute of Technical and Macromolecular Chemistry (ITMC), RWTH Aachen University, Worringerweg 2, 52074 Aachen, Germany, EU}

\author{Timon Kratzenberg}
\affiliation[IPC]
{Institute of Physical Chemistry, RWTH Aachen University, Landoltweg 2, 52074 Aachen, Germany, EU}

\author{Pablo Mota-Santiago}
\affiliation{Australian Synchrotron, ANSTO, Clayton, Victoria, Australia}
\alsoaffiliation[MAXIV]{MAX IV Laboratory, Lund University, P.O. Box 118, 22100 Lund, Sweden, EU}

\author{Andrea Scotti}
\affiliation[Lund]
{Division of Physical Chemistry, Lund University, SE-22100 Lund, Sweden}

\author{Andrij Pich}
\affiliation[DWI]
{DWI – Leibniz Institute for Interactive Materials, Forckenbeckstr. 50, 52074 Aachen, Germany, EU}
\alsoaffiliation[ITMC]
{Functional and Interactive Polymers, Institute of Technical and Macromolecular Chemistry (ITMC), RWTH Aachen University, Worringerweg 2, 52074 Aachen, Germany, EU}
\alsoaffiliation[AMIBM]
{Aachen Maastricht Institute for Biobased Materials (AMIBM), Maastricht University, Brightlands Chemelot Campus, Urmonderbaan 22, 6167 RD Geleen, The Netherlands, EU}

\author{Walter Richtering}
\affiliation[IPC]
{Institute of Physical Chemistry, RWTH Aachen University, Landoltweg 2, 52074 Aachen, Germany, EU}

\email{richtering@pc.rwth-aachen.de}

\title
{\large Unraveling the Mechanisms of Ultrasound-Induced Mechanical Degradation of Microgels: Effects of Mechanoresponsive Crosslinks, Softness, and Core-Shell Architecture}

\keywords{microgels, disulfide crosslinks, ultrasound, mechanochemical degradation, light scattering, small-angle X-ray scattering, atomic force microscopy}

\begin{document}
\newpage
\begin{abstract}
Ultrasound-induced degradation of soft polymeric colloids, like microgels, as well as a controlled drug release enabled by mechanoresponsive bonds, has recently attracted considerable attention. However, most examples in the literature focus primarily on the applications rather than examining the underlying mechanisms of the structural changes occurring in microgels due to cavitation – changes that are crucial for developing effective drug delivery systems.	
In this work, we provide a comprehensive view on how microgel structure governs the susceptibility to rupture and mass loss upon cavitation, investigating both conventional microgels containing mechanoresponsive disulfide bonds and more complex asymmetrically crosslinked core-shell microgels. 
By combining dynamic and static light scattering, small-angle X-ray scattering, and atomic force microscopy, we demonstrate that an interplay between mechanoresponsive crosslinks and the swelling degree determines the microgels susceptibility to ultrasound-induced damage. Our findings indicate that local stress from cavitation bubbles varies strongly within the microgel dispersion. The majority of microgels undergo gradual erosion at their periphery, resulting in smaller yet structurally intact particles over time, observable by light scattering and AFM. In contrast, microgels closer to a cavitation bubble can experience partial rupture or completely disintegrate, producing smaller, more polydisperse fragments, which contributes substantially to the overall mass loss observed. In the core–shell microgels with different crosslinkers in the core and shell, degradation occurs nearly uniformly across both regions, instead of selectively targeting the weaker part.
These observations highlight the complexity of the degradation dynamics as well as the similarity to processes seen in linear polymers and bulk hydrogels.
\end{abstract}

Ultrasound treatment is widely used in medicine for diagnostic imaging, local heating or tissue ablation~\cite{mason2011review}.
For imaging and most therapeutic applications, a low-intensity ultrasound in the MHz frequency range is used, which is harmless to the surrounding tissue, and its effect is limited to local heating~\cite{barnett1997}.
Conversely, high-intensity focused ultrasonic beams offer a non-invasive but efficient stimulus to trigger drug release in desired parts of the body and promote its internalization by cells~\cite{pitt2004review,yildiz2022review}.
This therapeutic approach, often termed sonopharmacology, relies on transforming the energy of the sound wave into a local mechanical force, similar to classical sonochemistry~\cite{suslick1990} and polymer mechanochemistry~\cite{ghanem2021review}.

In contrast to most medical applications, sonochemistry uses a low-frequency ($\geq20$~kHz) ultrasound of high intensity.
When such ultrasound waves pass through a liquid, they induce gas bubbles (cavities) that oscillate in phase with the wave and then spontaneously implode~\cite{lauterborn1985,akhatov2001,lauterborn2010review}.
This phenomenon, known as inertial cavitation, produces huge local shear or elongational strain in the vicinity of the bubble, with strain rates on the order of $10^6$-$10^7$~s$^{-1}$~\cite{nyborg1982,caruso2009review}.
Additionally, a collapsing bubble emits an intense shock wave from its center.
This wave travels at supersonic speeds and can have amplitudes as high as $10^4$-$10^5$~bar~\cite{pecha2000}. 
The mechanical forces resulting from inertial cavitation and shock waves are sufficient to break covalent bonds, destroy polymer molecules or drug carriers and release their cargo.

Designing an optimal mechanoresponsive drug carrier requires a thorough understanding of its activation/degradation process.
In the case of simple linear polymers, the basics of the degradation mechanism in ultrasonication experiments are fairly well understood~\cite{glynn1972,basedow1977chapter}.
Hence, the majority of recent research in the field of polymer mechanochemistry focuses on developing and incorporating mechanically responsive moieties -- mechanophores -- into polymers to achieve selective bond scission~\cite{berkowski2005, gostl2016,zhang2017,chen2020,muramatsu2021,ghanem2021review} and to trigger the release of small molecules~\cite{shi2020,huo2021,zou2022}.
However, the potential mechanoresponsive drug carriers, are often significantly larger and possess a more complex structure, which can have a dramatic influence on the mechanism and efficiency of mechanical degradation~\cite{zhang2015chapter}.
Examples of such materials include polyelectrolyte multilayer capsules~\cite{chen2017}, polymeric micelles~\cite{wu2017}, polymerosomes~\cite{rifaie2021}, composite hydrogels~\cite{epstein2010,fang2020} and microgels~\cite{di2017,zou2022}.

Microgels, which are crosslinked polymeric colloids swollen in a good solvent, represent arguably the most versatile prototype of a polymeric drug carrier.
They typically have a submicrometer size, low size polydispersity and a porous internal structure, all of which are well understood, can be easily controlled in synthesis and characterized using scattering or microscopy techniques~\cite{karg2019review,scheffold2020review}.
Furthermore, microgels can respond to external stimuli, like temperature~\cite{fernandez-barbero2002}, pH~\cite{fernandez-nieves2000}, or ionic strength~\cite{fernandez-nieves2001} by swelling and deswelling the polymer network.
The extent of swelling, which is controlled by the amount of crosslinker and solvent quality~\cite{lopez2017,braun2024}, defines the porosity of microgels allowing to load them with small molecules~\cite{nolan2006,braun2024a}, proteins~\cite{gawlitza2013a,sigolaeva2014} or nanoparticles~\cite{gawlitza2013,brasili2023}, and then release the loaded cargo on demand.
Due to the soft nature of microgels~\cite{voudouris2013,schulte2021,scotti2022review}, mechanical force can be used as an external trigger to selectively actuate or modify them, for example by ultrasound irradiation~\cite{schulte2022review}.

The application of a low-intensity ultrasound, especially in the MHz frequency range, does not break strong covalent crosslinks in microgels.
It can induce a deswelling of pNIPAm microgels~\cite{stock2024} or collapse of linear pNIPAm chains~\cite{rahimzadeh2021}.
Another study reported an increased swelling of microgels in response to mild ultrasonic treatment~\cite{venegas2013}.
Furthermore, absorption or scattering of mild ultrasonic waves by microgels can enhance the deformation of their embedding matrix, making the matrix also responsive to ultrasound~\cite{joshi2018}.
In contrast, application of a high-intensity ultrasound at lower frequency (20~kHz or similar) typically leads to irreversible mechanical degradation of microgels.
This effect was used in several studies to achieve on-demand release of encapsulated drugs~\cite{vukicevic2015,di2017,lei2020,kubota2021}.
Notably, microgels can only be degraded in their swollen state, while collapsed microgels are resistant to degradation by ultrasound~\cite{vukicevic2015,kharandiuk2022,he2023}. 
It was also shown that incorporation of mechanophores, such as disulfide or diselenide bonds, as crosslinkers or linkers for drug molecules can significantly facilitate microgel degradation~\cite{kharandiuk2022} or drug release~\cite{zou2022}, respectively. 
However, supramolecular crosslinks, such as catechin hydrogen bonds, can act as dynamic bonds and render microgels more resistant against mechanical forces~\cite{izak-nau2022}.

The majority of the aforementioned studies showed only the proof-of-principle results that microgels can be degraded by ultrasound.
Only a few works attempted to characterize the structural changes that occur in microgels as a result of cavitation.
\citeauthor{izak-nau2020} investigated poly($N$-vinylcaprolactam) (pVCL)-based microgels that were crosslinked with a dimethacrylate-functionalized mechanofluorophore.
Dynamic light scattering (DLS), cryo-transmission electron microscopy (cryo-TEM), and high-resolution magic-angle spinning nuclear magnetic resonance (HR-MAS NMR) were employed to demonstrate that the outer fuzzy shell of a microgel degrades first upon exposure to shearing forces.
This results in a more homogeneous internal structure and a lower swelling degree of the microgels~\cite{izak-nau2020}.
However, cryo-TEM may not always capture the full extent of structural changes in soft, highly-swollen particles, like microgels, because of contrast limitations.
Also, DLS does not provide direct structural information as it measures the average diffusion coefficient.
Alternatively, atomic force microscopy (AFM) can resolve many structural features of microgels~\cite{schulte2018,bochenek2019,schulte2019,schulte2021,nishizawa2019,nishizawa2023} and, therefore, is better suited to follow microgel degradation~\cite{schulte2022review}.
A few other studies showed cavitation-induced fragmentation of microgels into irregular debris~\cite{vukicevic2015,kharandiuk2022}.
However, to the best of our knowledge, the mechanism of this fragmentation remains unknown.

Here, we investigate the mechanism and efficiency of microgel degradation by applying high-intensity ultrasound.
We use dynamic light scattering (DLS), static light scattering (SLS), small-angle X-ray scattering (SAXS), and atomic force microscopy (AFM) to characterize the structural changes occurring in the microgels upon application of ultrasonication.
The combination of reciprocal- and real-space techniques allows us to detect even small structural modifications and, at the same time, overcome the limitations of increasing size polydispersity.

We examine both conventional covalently crosslinked microgels and those containing mechanoresponsive disulfide crosslinks to better understand how incorporated mechanophores influence the efficiency of degradation.
In addition, we investigate asymmetrically crosslinked core-shell microgels in which the disulfide crosslinks are located exclusively in either core or shell, hypothesizing that the mechanoresponsive part (core or shell) might be selectively degraded while the other part remains intact.

The presented findings reveal that mechanical degradation of microgels by ultrasound is more complex than it was considered in previous studies~\cite{izak-nau2020,kharandiuk2022,he2023}, both in terms of the influence of mechanophores and the underlying mechanism.

\newpage
\section{Experimental Section}

\subsection{Materials}
\textit{N,N´}-Bis(acryloyl)cystamine (BAC, $\geq95\%$, TCI), \textit{N,N´}-methylenbisacrylamide (BIS, $99\%$, Sigma-Aldrich), ammonium persulfate (APS, $\geq98\%$, Sigma-Aldrich), sodium dodecyl sulfate (SDS, $\geq99\%$, Sigma-Aldrich), and tetramethylethylenediamine (TEMED, $99\%$, Sigma-Aldrich) were used as received and without further purification. The monomer \textit{N}-isopropyl\-acryl\-amide (NIPAm, $>98\%$, TCI) was destabilized by recrystallization from \textit{n}-heptane.

\subsection{Syntheses}
All microgels were synthesized using precipitation polymerization in deionized water. As a crosslinker either BIS or BAC were used. The reactions were monitored and optimized using an in-line reaction calorimeter. This allowed us to monitor the real-time heat generation rate and the turbidity during the reaction. 
For quantification of the incorporated BAC content using Raman spectroscopy, homopolymers of pNIPAm and pBAC were synthesized in deionized water.
Table~S1 (Supporting Information) provides internal sample codes (from lab journal) for future reference.

\subsubsection{Synthesis of Conventional Microgels in an In-Line Reaction Calorimeter}
The in-line reaction calorimeter was used to optimize the microgel polymerizations. Depending on the initiation system used, different polymerization temperatures were chosen. For APS-initiated syntheses, a polymerization temperature of 70~$^{\circ}$C was used while for the APS/TEMED redox-initiated syntheses, a temperature of 50~$^{\circ}$C was used. The redox-initiated synthesis was chosen in order to avoid a thermal degradation of the disulfide crosslinker BAC as reported in the literature~\cite{gaulding2012}. In this work, we intended to keep the disulfide bond of the crosslinker intact for further degradation experiments applying ultrasound.
For online monitoring of the microgel formation process, syntheses were performed in an RC1e reactor from Mettler-Toledo (triple-walled, 0.5\;L), which is equipped with a downward-facing propeller stirrer, a turbidity sensor, and two temperature sensors. For the syntheses of the microgels, NIPAm, the surfactant SDS, and the crosslinker (BIS or BAC) with varying crosslinker content (1\;mol\% or 5\;mol\%) were dissolved in deionized water. In case of the APS/TEMED-initiated microgels, TEMED was added to the mixture using a micro liter pipette. The mixture was degassed (30\;min) using nitrogen, while stirred (300\;rpm), and heated to the desired reaction temperature. The initiator (APS) was added in solid form. The polymerization was terminated by cooling the reactor to 25~$^{\circ}$C after 1 to 3\;hours. The temperature inside the reactor, the heat flow of the reaction as well as the turbidity were measured over the course of the whole reaction. For measurements and data analysis, the software iControl RC1e 5.3 was used.
The exact amounts of chemicals used for the microgel syntheses performed in a reaction calorimeter are listed in Table~S2 (Supporting Information). 

\subsubsection{Syntheses of Conventional Microgels in Round-Bottom Flask}
For the synthesis of a pNIPAm microgel in a round-bottom flask, NIPAm, BAC or BIS, SDS, and TEMED (1\;mol\%) were dissolved in deionized water at room temperature and purged with nitrogen for 30\;min. Subsequently, under constant stirring (500\;rpm) the solution was heated up (50\;$^{\circ}$C) in an oil bath. To initiate the reaction, APS (1\;mol\%) was dissolved in deionized water (0.5\;mL) and added to the reaction solution. The polymerization was terminated after 1\;hour by slowly cooling down to room temperature. The molar percentage was calculated in correlation to the amount of NIPAm added. For purification of the microgel dispersion, it was dialyzed for at least 5\;days against deionized water (5\;L) with a cellulose membrane (MWCO 12–14\;kDa). The water was exchanged at least twice a day. After purification and lyophilization, the yield was determined gravimetrically.
The exact amounts of chemicals used for the synthesis of conventional pNIPAm-based microgels are listed in Table~S2 (Supporting Information).

\subsubsection{Synthesis of Core-Shell Microgels in Round-Bottom Flask}
The core-shell microgels were synthesized \textit{via} seeded precipitation polymerization in a round-bottom flask. NIPAm, TEMED, BAC or BIS, and purified core microgel solution (Tables~S3 and~S4, Supporting Information) were dissolved in deionized water (60\;mL) at room temperature in a round-bottom flask. The dispersion was purged with nitrogen for 30\;min. Subsequently, under constant stirring (500 rpm) the solution was heated (50\;$^{\circ}$C) using an oil bath. To initiate the reaction, APS was dissolved in deionized water (0.5\;mL), and added to the reaction solution. The polymerization was terminated after 1\;hour by slowly cooling down to room temperature. The molar percentage was calculated in correlation to the amount of NIPAm added. For purification of the microgel dispersion, it was dialyzed for at least 5\;days against deionized water (5\;L) with a cellulose membrane (MWCO 12–14\;kDa). The water was exchanged at least twice a day. After purification and lyophilization the yield was determined gravimetrically. 
The exact amounts of chemicals used for the synthesis of the core-shell microgels are listed in Table~S5 (Supporting Information).

\subsection{Raman Spectroscopy}
FT-Raman spectra were recorded with a Bruker RFS 100/S spectrometer. The samples were stimulated by a Nd:YAG laser with a wavelength of 1064\;nm and an output power of 200\;mW. Each sample was measured with 1000\;scans and covered a spectral range from 400 to 4000\;cm$^{-1}$ with a resolution of 4\;cm$^{-1}$. The baseline correction and the analysis of the recorded spectra were carried out with the software OPUS 4.0.\\
The quantification of the BAC content within the microgels was performed using calibration curves. Therefore, pNIPAm and pBAC were mixed in specific ratios and homogenized by dissolving both in methanol. The solvent was removed by evaporation at room temperature, and the solid samples were measured to obtain a calibration curve.

\subsection{Ultrasonication Experiments}
Sonochemical irradiation experiments were performed under an inert atmosphere on a Vibra-Cell ultrasonic processor VCX500 (Sonics \& Materials) in a three-neck Suslick vessel (Zinsser Analytic). For each degradation experiment, the pristine microgels were diluted (0.5\;mg\;mL${^{-1}}$), filled into the Suslick vessel, which was sealed with septa, placed in an ice bath, and then degassed with nitrogen for 10\;min. The microgel dispersions were sonicated with a 13\;mm probe (maximum displacement amplitude of $A_{\textrm{max}}=115$\;µm) for different periods of time between 5 and 180\;min in a pulsed ultrasonication mode (1\;s “on” and 2\;s “off”, i.e. 33\% duty cycle). The frequency was $f=20$\;kHz, and the amplitude was kept at 30\%.
These settings correspond to the nominal ultrasound intensity of $I_{\mathrm{nom}}=2\pi^2f^2A^2\rho c \approx1.5\cdot10^8$\;W/m$^2$ and nominal acoustic pressure of $P_{\mathrm{nom}}=2\pi fA\rho c \approx 64$\;bar \cite{basedow1977chapter}.

\subsection{Dynamic and Static Light Scattering}
Multi-angle dynamic light scattering (DLS) was used to determine the hydrodynamic radius of the microgels, $R_{\textrm{h}}$.
The measurements were performed using an ALV instrument equipped with a HeNe laser (wavelength in vacuum  $\lambda_0 = 633$~nm) and a digital correlator.
The measurements were performed at $T=20\;^{\circ}$C in double-distilled water (refractive index $n(\lambda_0)=1.332$) filtered through a 0.2~µm syringe filter.
The temperature was controlled using a toluene bath to match the refractive index of the glass vial and an external bath circulator.
The scattering intensity was measured at scattering vectors $q=4\pi n/\lambda_0 \sin{\theta/2}$, where the scattering angle $\theta$ was changed between 30$^{\circ}$ to 110$^{\circ}$ degrees in 5$^{\circ}$ or 10$^{\circ}$ steps.
The time of acquisition was 60~s per angle.

Static light scattering (SLS) measurements were performed in double-distilled water filtered through a 0.2~µm syringe filter using an SLS-Systemtechnik GmbH instrument equipped with a blue laser ($\lambda_0 = 407$~nm), a far-red laser  ($\lambda_0 = 819$~nm) and an index-matched toluene bath.
The scattering intensity was measured at $T=20\;^{\circ}$C as a function of the scattering vector $q$ by changing the scattering angle $\theta$ between 20$^{\circ}$ and 150$^{\circ}$ with a 2$^{\circ}$ step.

\subsection{Small-Angle X-Ray Scattering}
The small-angle X-ray scattering (SAXS) experiments were performed at the CoSAXS beamline at the 3 GeV synchrotron ring of the MAX-IV Laboratory (Lund, Sweden)~\cite{plivelic2019}. 
The $q$-range between $1\cdot10^{-2}$ and $0.7$~nm$^{-1}$ was covered on the CoSAXS using a sample-to-detector distance of 10~m and the X-ray beam energy of $E = 12.4$~keV.
The Eiger2 4M SAXS detector with pixel size of 75~$\mu$m x 75~$\mu$m was used. 
The data were reduced using the Python scripts used at the beamline.

The scattering intensity $I(q)$ was fitted with the form factor model of a fuzzy-sphere for regular microgels~\cite{stieger2004} (SAXS data) and a fuzzy core-shell for the core-shell microgels~\cite{berndt2005,dubbert2014} (SLS data) using a custom Matlab-based software \textit{FitIt!}~\cite{virtanen2016}.
In the fuzzy-sphere form factor model, it is assumed that a particle of radius $R$ consists of a homogeneous core of radius $R_{\mathrm{c}}$ and a fuzzy shell of width $2\sigma$, such that $R = R_{\mathrm{c}} + 2\sigma$.
In the fuzzy core-shell form factor model, a particle of radius $R$ is assumed to have a homogeneous core of radius $R_{\mathrm{c}}$, an interpenetrating layer between the core and the shell of the length $2\sigma_{\mathrm{in}}$, a homogeneous shell of length $w$ and a fuzzy outer surface of the length $2\sigma_{\mathrm{out}}$.
The total particle radius is given by a sum of the above parts, $R = R_{\mathrm{c}} + 2\sigma_{\mathrm{in}} + w + 2\sigma_{\mathrm{out}}$.
The size polydispersity in both models is accounted for by convolving the form factor with a Gaussian distribution of the total radius $R$~\cite{stieger2004,berndt2005}.
An additional Lorentzian term with an average mesh size $\xi$ can be added to account for scattering from density inhomogeneities of the polymer network at high $q$-values (irrelevant for SLS data)~\cite{stieger2004,fernandez-barbero2002}.

\subsection{Atomic Force Microscopy}

The samples for dry-state atomic force microscopy (AFM) measurements were prepared \textit{via} dip coating on silicon substrates (ca. 1.5x1.5~cm).
The substrates were cleaned in isopropanol using an ultrasound bath and activated using a UV Ozone Cleaner (NanoBioAnalytics UVC-1014) prior to deposition.
The microgel dispersions at a concentration of 3.5 mg mL$^{-1}$ were mixed with isopropanol in a 1:4 ratio (v/v), and between 30 and 120~µL of the mixture was deposited on a custom glass trough filled with Milli-Q water (surface area $A_{\textrm{tot}}=50.27$~cm$^2$).
The substrates were lifted out of the trough at a speed of $0.67~\si{\milli\m\per\s}$ following 25 minutes of equilibration.

The AFM images were recorded using a Nanosurf NaioAFM instrument and a BudgetSensors TAP190Al-G probe (cantilever) with a spring constant of $48~\si{\N\per\m}$, a resonant frequency of $190~\si{\kilo\Hz}$ and a nominal tip radius of 10~nm.
The tapping mode of the instrument was used with a vibrational amplitude of $\approx230$~mV.
The AFM images were corrected using the Gwyddion software (version 2.61)~\cite{nevcas2012}.
A custom Matlab script (MathWorks, 2022b) was employed for the measurement of the contact radii $R_\mathrm{cont}$ and the AFM height profiles.
The script automatically detects the microgel centers based on the height image data and assigns a region of interest (ROI) to the corresponding phase image.
The ROI is then manually adjusted to measure $R_\mathrm{cont}$ assuming a circular geometry of the individual microgels spread at the interface.
The microgel detection is based on the particle tracking algorithm described by Crocker \textit{et al}.\cite{Crocker.1996, bochenek2019}
For each microgel, two height profiles are generated by linear interpolation of the height data around the microgel center parallel and orthogonal to the fast scan direction of the AFM image, respectively.

For each microgel sample corresponding to a different ultrasonication time $t$, between 5 and 16 AFM images were recorded, so that between 52 and 403 individual microgels were analyzed by the script.
The statistical analysis was performed with the one-way ANOVA test and Tukey-Kramer post-hoc pairwise comparison using built-in functions in Matlab (MathWorks, 2022b): \textit{anova1} and \textit{multcompare}, significance level 0.05.
\newpage

\section{Results}

\subsection{Microgel Synthesis and Ultrasonication}

The microgels used in this study were synthesized by precipitation polymerization of \linebreak \textit{N}-isopropyl\-acryl\-amide (NIPAm) in demineralized water~\cite{pelton1986} using either a conventional cross\-linker \textit{N,N'}-methylene\-bis\-acryl\-amide (BIS) or a mechanoresponsive disulfide crosslinker \linebreak \textit{N,N'}-bis\-(acryloyl)\-cystamine (BAC).
The microgels containing a single type of crosslinker and synthesized in a single polymerization step will be referred to as \textit{conventional} microgels, Table~\ref{tbl:Rh}.
Additionally, asymmetrically crosslinked \textit{core-shell} microgels were synthesized with a different crosslinker in the core and shell using a seeded precipitation polymerization with the conventional microgels acting as seeds~\cite{berndt2003}.
The hydrodynamic radii, $R_{\textrm{h}}$, of all the obtained microgels in the swollen state ($T=20\;^{\circ}$C, below the VPT temperature) and in the collapsed state ($T=40\;^{\circ}$C, above the VPT temperature) are shown in Table~\ref{tbl:Rh}.
We calculate the swelling ratio of the microgels as the quotient of the swollen and collapsed hydrodynamic radius, $\alpha = R_{\textrm{h}}(20\;^{\circ}\textrm{C})/R_{\textrm{h}}(40\;^{\circ}\textrm{C})$.
The swelling ratio is used as a measure of the softness of a microgel – the higher the swelling ratio, the softer the microgel~\cite{scotti2022review}.
Microgels with a higher amount of crosslinker (5~mol\% \textit{vs.} 1~mol\%) have a lower swelling ratio making them effectively harder, as expected~\cite{houston2022}, Table~\ref{tbl:Rh}.
Interestingly, the microgels with BAC crosslinker have smaller values of $\alpha$ than the microgels with the same nominal amount of BIS crosslinker.
This observation is unexpected but it can be explained by a significantly higher hydrophobicity of the BAC molecules, as well as by side-reactions of the BAC crosslinker that might occur during the synthesis~\cite{gaulding2012}.
A detailed discussion of the difference in swelling of the microgels, supported by in-line reaction calorimetry and Raman spectroscopy data, is given in Sections~S2-S4 and Figures~S1-S3 (Supporting Information).

\begin{table}[ht!]
  \caption{Characteristics of the investigated microgels, including their hydrodynamic radii in swollen state $R_{\textrm{h}}(20\;^{\circ}\textrm{C})$, in the collapsed state $R_{\textrm{h}}(40\;^{\circ}\textrm{C})$, and the swelling ratios $\alpha$.}
  \label{tbl:Rh}
  \resizebox{\textwidth}{!}{%
  \begin{tabular}{cccccc}
    \hline
    Name & Type of microgel & Crosslinker (core/shell) & $R_{\textrm{h}}(20\;^{\circ}\textrm{C})$ [nm] & $R_{\textrm{h}}(40\;^{\circ}\textrm{C})$ [nm] & $\alpha$\\
    \hline
    BIS-1 & conventional & 1~mol\% BIS & $142\pm2$ & $50.5\pm0.4$ & $2.82\pm0.05$ \\
    BIS-5 & conventional & 5~mol\% BIS & $140\pm3$ & $66.1\pm0.5$ & $2.11\pm0.04$ \\
    BAC-1 & conventional & 1~mol\% BAC & $127.7\pm0.7$ & $66.4\pm0.5$ & $1.92\pm0.02$ \\
    BAC-5 & conventional & 5~mol\% BAC & $103.3\pm0.5$ & $69.0\pm0.8$ & $1.50\pm0.02$ \\
    CS-A  & core-shell & 1~mol\% BAC/1~mol\% BIS & $380\pm8$ & $156\pm2$ & $2.44\pm0.06$ \\
    CS-B  & core-shell & 1~mol\% BIS/1~mol\% BAC & $318\pm12$ & $151\pm3$ & $2.11\pm0.09$ \\
    \hline
  \end{tabular}}
\end{table}

Mechanical degradation of the microgels was achieved by irradiating them under a nitrogen atmosphere with 20~kHz ultrasound using a sonotrode, Scheme~\ref{scheme_1}.
An ice bath was used to prevent the heating of the irradiated samples and to ensure that the microgels remained in the swollen state.
After the ultrasonication, any cleaved polymer chains and network fragments were removed by centrifugation at 50,000 rpm and only the purified microgels were analyzed further.
We performed a trial DLS measurement of the supernatant, which showed that the intermediate scattering function, $f(q,\tau)$, did not follow a single-exponential course, and the obtained decay rates, $\Gamma(q)$, did not strictly obey the $q^2$ scaling.
Therefore, a light scattering characterization of the supernatant requires its separation into relatively monodisperse fractions, which is out of the scope of this manuscript and will be addressed in a separate study.

\begin{scheme}[htpb!]
    \centering
    \includegraphics[width=\linewidth]{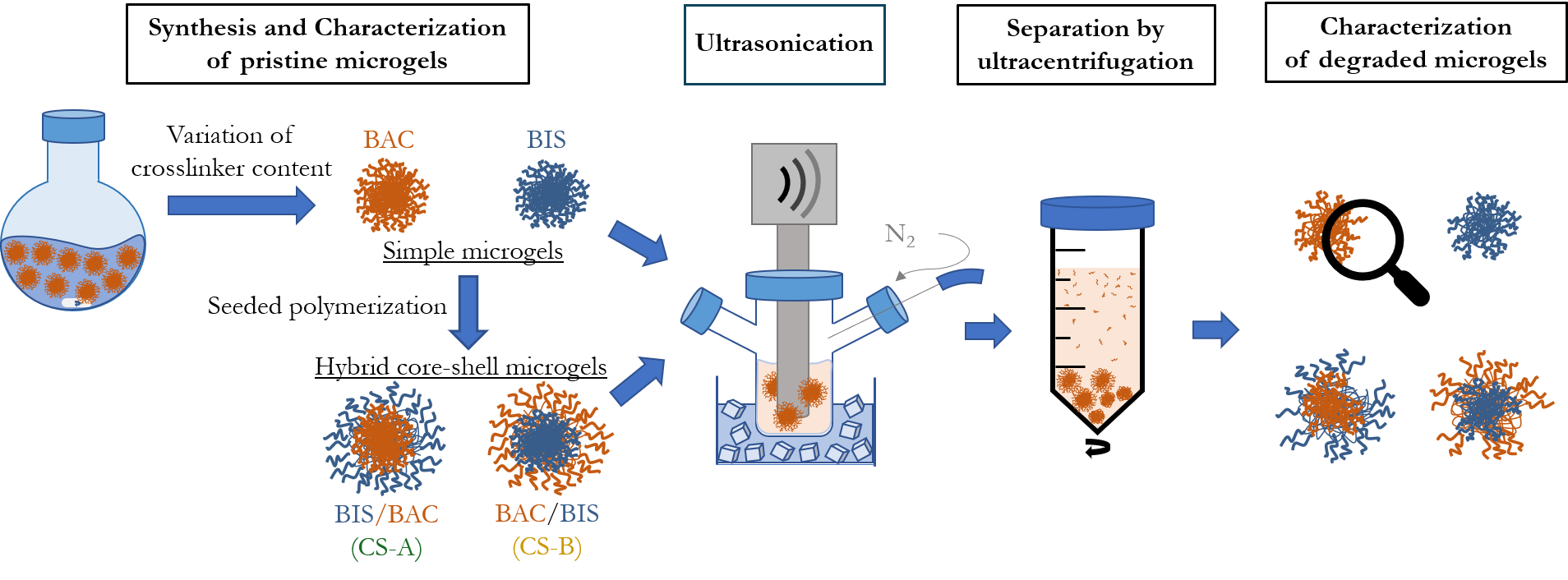}
    \caption{Schematic of the experimental approach used in this study: after ultrasonication, microgels are separated from any cleaved chains and fragments by ultracentrifugation and then analyzed using various scattering methods and atomic force microscopy.}
    \label{scheme_1}
\end{scheme}

\subsection{Degradation of Microgels Observed by DLS}

\begin{figure*}[ht!]
    \centering
    \includegraphics[width=\linewidth]{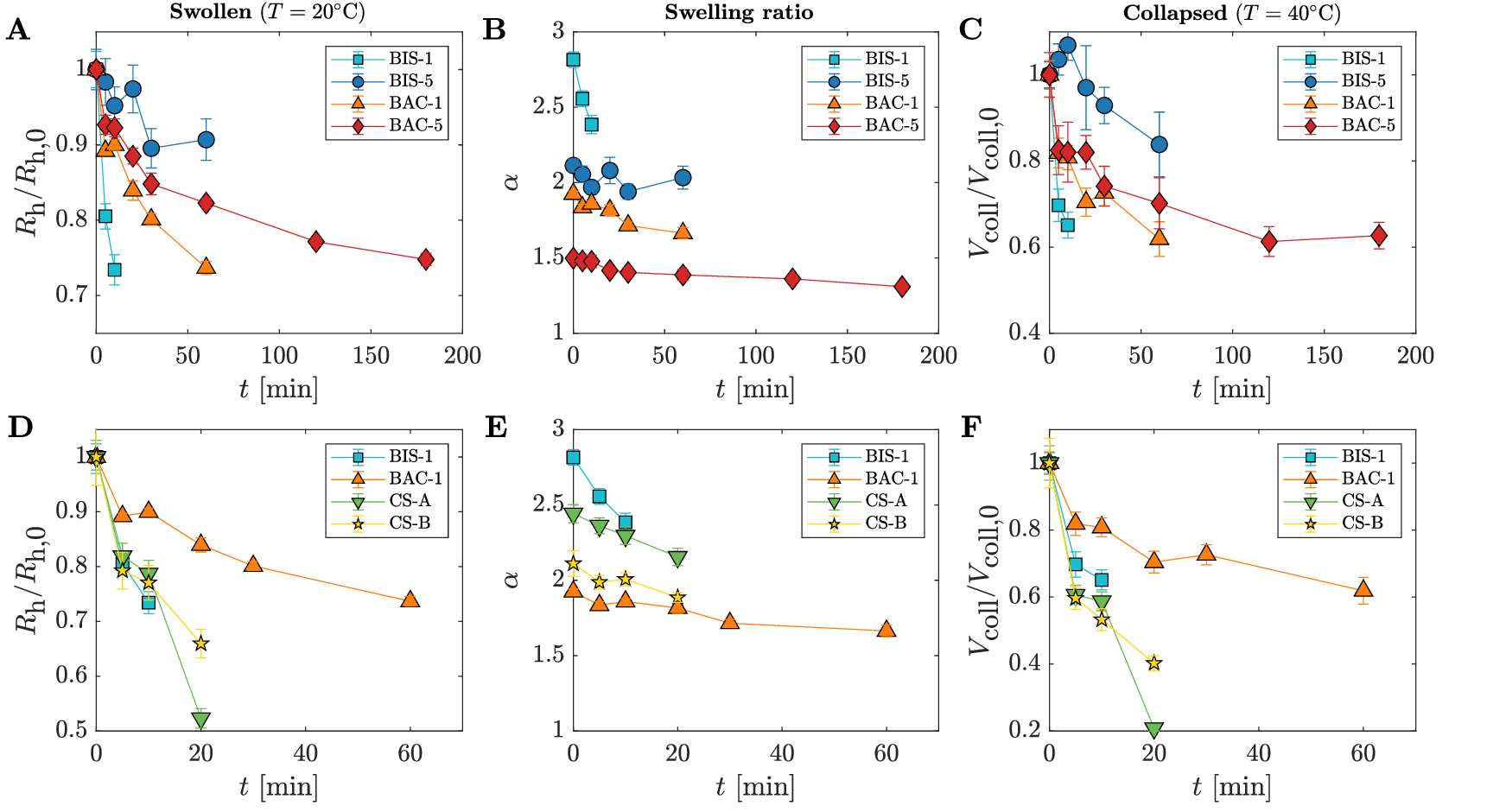}
    \caption{(\textbf{A,D}) Normalized hydrodynamic radii of the conventional microgels (\textbf{A}) and core-shell microgels (\textbf{D}) in the swollen state ($T=20\;^{\circ}$C), $R_{\textrm{h}}/R_{\textrm{h},0}$, after different ultrasonication times $t$. (\textbf{B,E}) Swelling ratio $\alpha$ of the conventional microgels (\textbf{B})  and core-shell microgels (\textbf{E}) after different ultrasonication times $t$. (\textbf{C,F}) Normalized volume of the conventional microgels (\textbf{C}) and core-shell microgels (\textbf{F}) in the collapsed state ($T=40\;^{\circ}$C), $V_{\textrm{coll}}/V_{\textrm{coll,0}}$, after different ultrasonication times $t$.}
    \label{fig:DLS}
\end{figure*}

First, we measure the decrease of the microgel size due to the ultrasonic irradiation using DLS.
This is commonly done in the literature to assess the stability of microgels against mechanical degradation~\cite{izak-nau2020,izak-nau2022,kharandiuk2022,zou2022,he2023}.
Figure~\ref{fig:DLS}A shows the hydrodynamic radii of the microgels measured by multi-angle DLS at $T=20\;^{\circ}$C after different ultrasonication times $t$, normalized by the hydrodynamic radius of the pristine microgels, $R_{\textrm{h},0}$.
The radii of microgels in the collapsed state ($T=40\;^{\circ}$C) are shown in Figure~S5A and their change with $t$ is qualitatively similar.
The hydrodynamic radii decrease gradually for all microgels but the rate of the decrease depends significantly on the amount and type of crosslinker.
Generally, microgels with fewer crosslinks degrade faster, for example \mbox{BIS-1} (cyan squares) \textit{vs.} \mbox{BIS-5} (blue circles) or \mbox{BAC-1} (orange triangles) \textit{vs.} \mbox{BAC-5} (red diamonds), in agreement with other studies~\cite{izak-nau2020,izak-nau2022}.
When comparing \mbox{BIS-5} (blue circles) and \mbox{BAC-5} microgels (red diamonds), it can be seen that the presence of mechanoresponsive bonds makes the microgels more degradable by ultrasound, also as expected~\cite{kharandiuk2022}.
In contrast, a comparison of \mbox{BIS-1} microgels (light blue squares) and \mbox{BAC-1} microgels (orange triangles) gives the opposite result: in this case BIS microgels are more susceptible to mechanical degradation.
To reconcile this discrepancy, we consider that BAC and BIS microgels have a different swelling degree, $\alpha$, Table~\ref{tbl:Rh}: \mbox{BIS-1} microgels have a lower value of $\alpha$ than \mbox{BAC-1} microgels.
Indeed, strong swelling is known to introduce internal stresses in polymer materials making them more susceptible to mechanical failure~\cite{metze2023}.
The effect of swelling dominates over the effect of the labile (disulfide) bonds for \mbox{BIS-1} \textit{vs.} \mbox{BAC-1} microgels.
The difference in swelling might also contribute to the small difference in mechanical stability between \mbox{BAC-5} and \mbox{BIS-5} microgels.
Therefore, the resistance of microgels to mechanical degradation results from a complex interplay between the presence of mechanoresponsive bonds, the amount of crosslinker and the swelling degree.
Over the course of ultrasonic degradation, the swelling degree of all microgels decreases in agreement with previous work~\cite{izak-nau2020}, Figure~\ref{fig:DLS}B.
This indicates that microgels should become more mechanically-resistant and degrade slower over time, which is in line with the data in Figure~\ref{fig:DLS}A.

The extent of degradation of the microgels can be quantified by their relative mass loss: the decrease of the molecular weight $M_{\mathrm{w}}$.
Since the size of a microgel in the collapsed state does not depend on the degree of crosslinking or internal structure~\cite{lopez2017}, the volume of a collapsed microgel obtained by DLS, $V_{\textrm{coll}}$, provides a good measure of its molecular weight, $M_{\mathrm{w}}$:
\begin{equation}
    M_{\mathrm{w}}\propto V_{\textrm{coll}}=\frac{4}{3}\pi R_{\textrm{h}}^3(40\;^{\circ}C).
\end{equation}

Figure~\ref{fig:DLS}C shows the course of the normalized volume of a collapsed microgel $V_{\textrm{coll}}/V_{\textrm{coll},0}$ with increasing ultrasonication time.
The data for all microgels do not follow a linear trend, indicating that degradation becomes slower with increasing $t$.
Surprisingly, the value of $V_{\textrm{coll}}/V_{\textrm{coll},0}$ for \mbox{BAC-5} microgels tends to a plateau at $t>100$~min.
In addition, \mbox{BIS-1} microgels after 10~min and \mbox{BAC-1} microgels after 60~min showed a very low scattering intensity (not shown), such that sufficiently good data could not be obtained at larger ultrasonication times for these microgels.
Yet, the relative decrease of molecular weight (collapsed volume) is unexpectedly small, less than $40\%$.

Further unexpected observations can be found in the DLS data for the core-shell microgels.
Figure~\ref{fig:DLS}D shows the normalized hydrodynamic radii \textit{vs.} ultrasonication time $t$ for the microgels with BAC crosslinker in the core and BIS in the shell, \mbox{CS-A} (green downward triangles), and \textit{vice versa} BIS in the core and BAC in the shell, \mbox{CS-B} (yellow stars).
Because of the presence of both BIS and BAC crosslinkers, as well as intermediate swelling degree (Table~\ref{tbl:Rh} and Figure~\ref{fig:DLS}E), the core-shell microgels are expected to have a stability intermediate between the two conventional microgels with the same amount of crosslinker: \mbox{BIS-1} (cyan squares) and \mbox{BAC-1} (orange downward triangles).
Yet, both \mbox{CS-A} and \mbox{CS-B} decrease their relative hydrodynamic radius approximately as fast as \mbox{BIS-1} microgels.
The relative molecular weight (collapsed volume) of the core-shell microgels decreases even faster than for \mbox{BIS-1} microgels, Figure~\ref{fig:DLS}F.
These observations indicate that the microgel degradation cannot be understood based solely on DLS data, but a more detailed structural characterization is needed.

\subsection{Structural Changes of Microgels Observed by Static Scattering}

\begin{figure}[ht!]
    \centering
    \includegraphics[width=\linewidth]{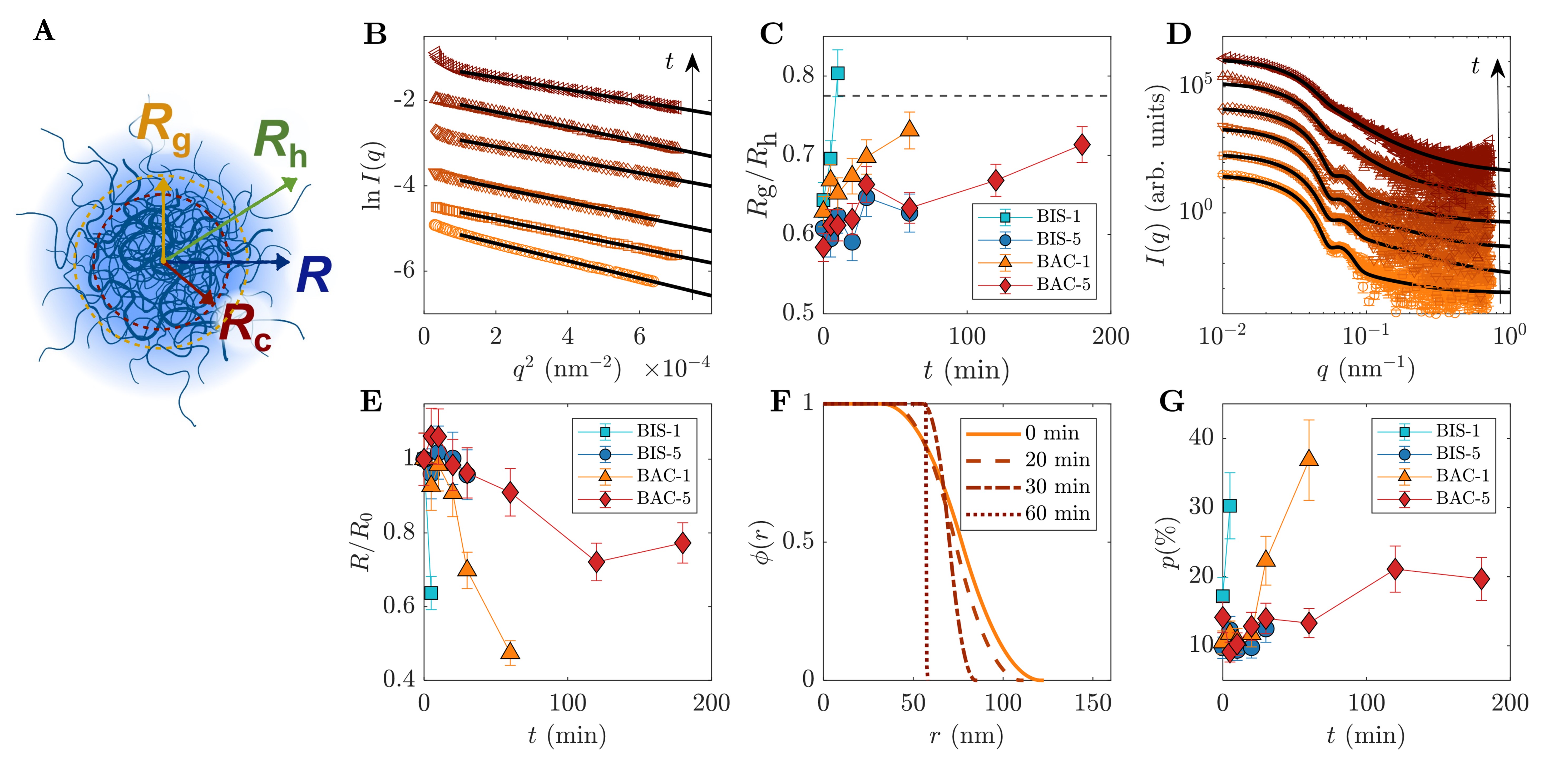}
    \caption{(\textbf{A}) Schematic of the different parameters describing microgel size and internal structure: the hydrodynamic radius $R_{\textrm{h}}$, the radius of gyration $R_{\textrm{g}}$, the total radius $R$ and radius of the core $R_{\textrm{c}}$ from the form factor fits. (\textbf{B}) Guinier plots of SLS intensity, $\ln{I(q)}$ \textit{vs.} scattering vector $q$, for \mbox{BAC-1} microgels after 0~min, 5~min, 10~min, 20~min, 30~min and 60~min ultrasonication (from bottom to top). Black lines correspond to the fits with Equation~\ref{eq:Guinier}. (\textbf{C}) The $R_{\textrm{g}}/R_{\textrm{h}}$ ratios for microgels \textit{vs.} ultrasonication time $t$. The dashed line correspond to the value for hard spheres, $R_{\textrm{g}}/R_{\textrm{h}}=0.778$. (\textbf{D}) SAXS intensities $I(q)$ \textit{vs.} scattering vector $q$ for \mbox{BAC-1} microgels after 0~min, 5~min, 10~min, 20~min, 30~min and 60~min ultrasonication (from bottom to top). Black lines correspond to the fits using the fuzzy-sphere model. (\textbf{E}) Normalized radii of the microgels $R/R_0$ obtained from the SAXS fits \textit{vs.} $t$. (\textbf{F}) Radial profiles of relative polymer density, $\phi(r)$, for \mbox{BAC-1} microgels, obtained from the SAXS fits. (\textbf{G}) Polydispersities of the microgels $p$, obtained from the SAXS fits \textit{vs.} $t$.}
    \label{fig:SAXS}
\end{figure}

The changes to the internal structure of microgels as a result of their mechanical degradation were first investigated using static scattering methods: SLS and SAXS.
These methods provide ensemble-average information and, because of the low polydispersity of the studied microgels, are highly sensitive to small structural changes.
SLS is used for the conventional BIS- or BAC-crosslinked microgels to gain a model-free structural information at the length scale of the whole particle (low values of the scattering vector $q$).
To fit the data, we use the Guinier approximation~\cite{glatter2018}:
\begin{equation}
    \ln{I(q)} = \ln{I_0} - \frac{R_{\mathrm{g}}^2}{3}q^2,
    \label{eq:Guinier}
\end{equation}
where $I_0$ is the forward scattering intensity and $R_{\textrm{g}}$ the average radius of gyration – a measure of radial distribution of mass in a particle.
This effective radius is always smaller than $R_{\textrm{h}}$, as shown schematically in Figure~\ref{fig:SAXS}A.
Figure~\ref{fig:SAXS}B shows the examples of Guinier plots, $\ln{I(q)}$ \textit{vs.} $q^2$, along with fits (black solid lines) for \mbox{BAC-1} microgels with increasing ultrasonication time (from bottom to top).
SLS data for other microgels and the $R_{\textrm{g}}$ values obtained from the fits are shown in Figure~S6 and Figure~S7 (Supporting Information).
The $R_{\textrm{g}}$ values also decrease with increasing sonication time $t$ for all microgels, but slower compared to $R_{\textrm{h}}$ values.
We can quantify this difference by calculating the quotient of the radius of gyration and hydrodynamic radius, $\rho = R_{\textrm{g}}/R_{\textrm{h}}$.
This value is a sensitive structural parameter related to the fuzziness of a particle~\cite{senff2000,pich2006,clara2012,elancheliyan2022}.
Figure~\ref{fig:SAXS}C shows the $\rho$-parameter \textit{vs.} $t$ for the investigated conventional microgels.
Before ultrasonic degradation, $\rho\approx0.6-0.65$ for all studied microgels, which is compatible with a fuzzy sphere structure of pNIPAm microgels~\cite{senff2000,elancheliyan2022}.
With increasing ultrasonication time, $\rho$ gradually increases and this increase proceeds faster for less mechanically stable microgels.
The increase of the structural parameter $\rho$ can be explained by removal of the dangling or loosely-crosslinked polymer chains at the microgel periphery, as discussed in previous studies~\cite{izak-nau2020,izak-nau2022,kharandiuk2022}.
Nevertheless, within the investigated ultrasonication times the values of $\rho$ only reach the limit of homogeneous spheres, $\rho_{\textrm{HS}}=0.778$~\cite{senff2000} (Figure~\ref{fig:SAXS}C, dashed line), for \mbox{BIS-1} microgels, which are the weakest.
We note that at this point the scattering signal is quite low for \mbox{BIS-1} microgels and the resulting $R_{\textrm{h}}$ is prone to error.
All other microgels do not reach the homogeneous-sphere limit, which means they retain a fuzzy structure to some extent for the whole duration of the ultrasound application.
Thus, the structural parameter $\rho$ allows quantifying the loss of dangling chains in the microgel periphery during ultrasonication that has only been qualitatively reported so far~\cite{izak-nau2020}.

Next, we perform SAXS measurements of the microgels in dilute suspension ($T=20\;^{\circ}$C, swollen state) before and after ultrasonic degradation.
This allows us to analyze changes in the full form factor $P(q)$ of the microgels, which describes their internal structure, by fitting the data to an appropriate model. 
Figure~\ref{fig:SAXS}D shows the SAXS intensities of \mbox{\mbox{BAC-1}} microgels with increasing ultrasonication times (from bottom to top).
The shoulder at $q\sim7\cdot10^{-2}$~nm$^{-1}$ shifts slightly to higher $q$-values and the minima become more smeared with increasing ultrasonication time $t$.
The data for other microgels are shown in 
Figure~S8 (Supporting Information).
The scattering intensities are fitted with the form factor model of a fuzzy sphere, which has been shown to accurately describe the internal structure of pNIPAm microgels~\cite{stieger2004,gasser2014,mohanty2017}, Figure~\ref{fig:SAXS}D (black solid lines).
The model provides the following parameters that describe the fuzzy-sphere structure: the total microgel radius $R$, the radius of the highly-crosslinked core $R_{\mathrm{c}}$, the width of the fuzzy shell $2\sigma$, and the size polydispersity $p$.
All the obtained fit parameters are given in Table~S8 (Supporting Information).
The radii of microgels $R$ after different ultrasonication times, normalized by the value before ultrasonication $R_0$, are shown in Figure~\ref{fig:SAXS}E.
In contrast to the hydrodynamic radii $R_{\mathrm{h}}$, the determined $R$ corresponds to the entire crosslinked network of a microgel including both core and fuzzy shell, but it excludes the dangling chains at the microgel periphery~\cite{senff2000}.
The difference between $R$ and $R_{\mathrm{h}}$ is shown schematically in Figure~\ref{fig:SAXS}A.

In the first 20~min of ultrasonication,  the $R/R_0$ values remain almost constant and close to 1 for all microgels except the \mbox{BIS-1}.
This is in stark contrast to the $R_{\textrm{h}}/R_{\textrm{h},0}$ values from DLS and the $\rho$-parameter from SLS/DLS, which both experience the strongest change within the first 20-30~min.
Therefore, we can conclude that during the initial degradation period only very loose dangling chains (visible mostly in DLS) are removed from the microgels, whereas most of the polymer network (visible in SAXS) remains relatively intact.
Here, the \mbox{BIS-1} microgels are an exception: they are completely degraded within 20~min under the experimental conditions because of their extreme softness.
The fuzziness of the microgels also hardly changes during the initial degradation period.
This is shown in Figure~\ref{fig:SAXS}F for \mbox{BAC-1} microgels using the radial distributions of relative polymer density, $\phi(r)$, obtained from the fits: solid line (0~min) \textit{vs.} dashed line (20~min).
Despite the decrease of $R_{\mathrm{h}}$ and increase of the $\rho$-parameter, microgels remain fuzzy and retain most of their shell.

After a certain ``critical'' ultrasonication time, which depends on the mechanical stability of the microgels, the polydispersity obtained from the SAXS fits $p$ increases dramatically, Figure~\ref{fig:SAXS}G.
For \mbox{BIS-1} microgels this happens already after 5~min of ultrasonication, for \mbox{BAC-1} microgels after 30 min and for \mbox{BAC-5} microgels after 120~min.
The increase of polydispersity means that microgel degradation proceeds non-uniformly resulting in a broad distribution of particle sizes or shapes.
Interestingly, the increase of polydispersity $p$ coincides with the strong decrease of the radii from the fit, $R/R_0$.
This means that removal of large parts of the polymer network inevitably leads to heterogeneous particles.
During this late degradation period, the fuzziness of \mbox{BAC-1} microgels decreases and the microgels eventually become homogeneous, Figure~\ref{fig:SAXS}F (dashed-dotted and dotted lines).
However, the large polydispersity at 30 and 60~min ($p>20\%$) smears the data significantly, so the size and internal structure extracted from these fits are subject to large uncertainties and should be treated with caution.
In contrast, \mbox{BAC-5} microgels, which are more resistant to mechanical degradation, remain fuzzy even after 180~min ultrasonication, Figure~S8C and Table~S8 (Supporting Information).
When microgels are completely degraded, the SAXS scattering profiles become featureless, and a unique meaningful fit is not possible, see Figure~S8A (Supporting Information).
It is important to note that the samples investigated by SAXS were first purified by ultracentrifugation.
This preparatory step most likely removed many of the smaller fragments formed during the ultrasonic treatment.
As a result, the measured fraction contains mostly larger, relatively uniform particles.
This has a particularly significant impact on the reported values for polydispersity, which may not fully capture the true extent of degradation-induced heterogeneity.

\begin{figure*}[ht!]
     \centering
     \includegraphics[width=\linewidth]{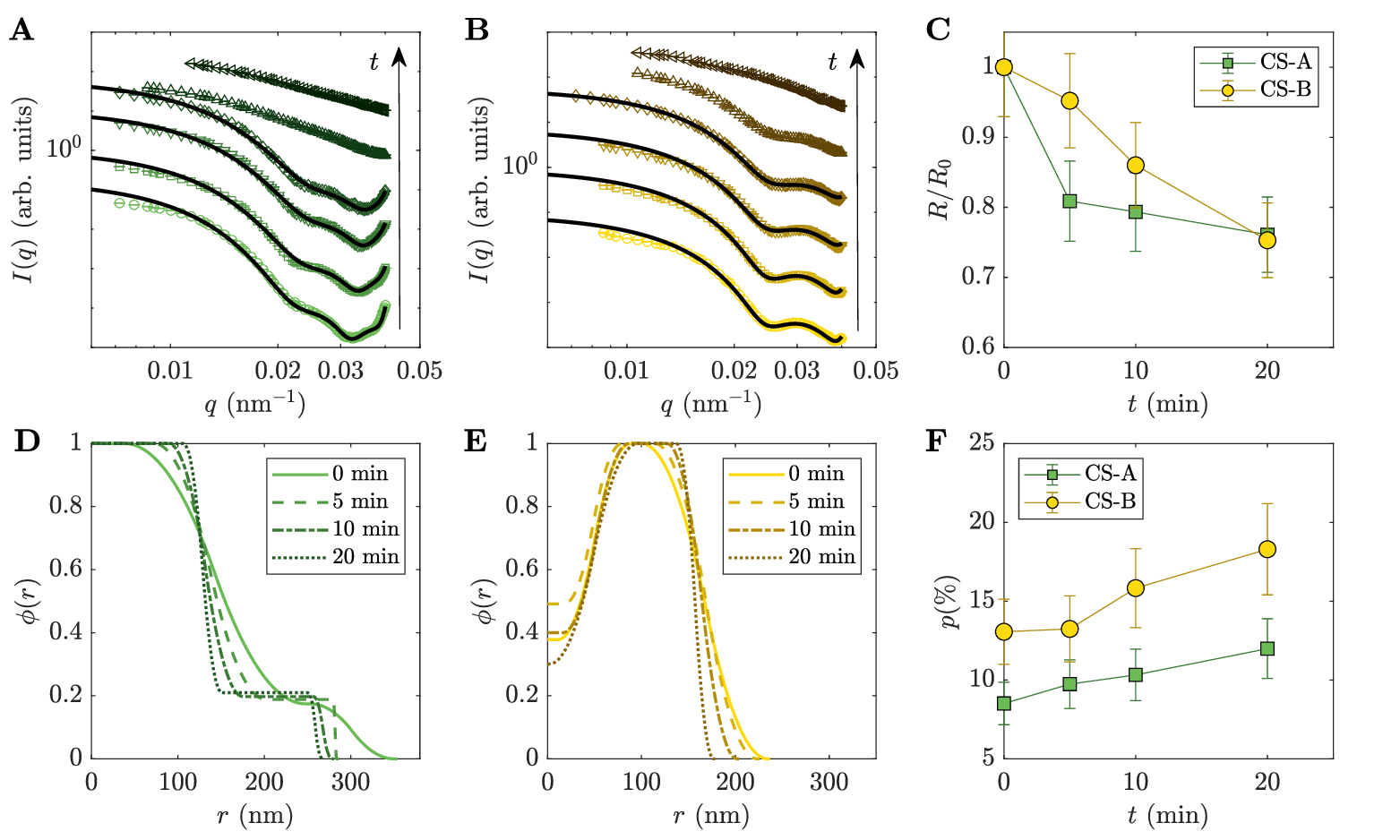}
     \caption{(\textbf{A}) SLS intensities $I(q)$ \textit{vs.} $q$ for \mbox{CS-A} microgels after 0~min, 5~min, 10~min, 20~min, 30~min and 60~min ultrasonication (from bottom to top). Black lines correspond to the fits using the fuzzy core-shell model. (\textbf{B}) SLS intensities $I(q)$ \textit{vs.} $q$ for \mbox{CS-B} microgels after 0~min, 5~min, 10~min, 20~min, 30~min and 60~min ultrasonication (from bottom to top). Black lines correspond to the fits using the fuzzy core-shell model. (\textbf{C}) Normalized radii of the core-shell microgels $R/R_0$ obtained from the SLS fits \textit{vs.} $t$. (\textbf{D}) Radial profiles of relative polymer density for \mbox{CS-A} microgels, $\phi(r)$, obtained from the SLS fits. (\textbf{E}) Radial profiles of relative polymer density for \mbox{CS-B} microgels, $\phi(r)$, obtained from the SLS fits.(\textbf{F}) Polydispersities of the core-shell microgels, $p$, obtained from the SLS fits \textit{vs.} $t$.}
     \label{fig:SLS_CS}
\end{figure*}

Since the \mbox{BIS-1} microgels are degraded faster than \mbox{BAC-1} microgels, we hypothesized that in the core-shell microgels, \mbox{CS-A} and \mbox{CS-B}, the ``weaker'' part (core or shell, crosslinked with BIS) is degraded first, followed by the ``stronger'' part (shell or core, crosslinked with BAC, respectively).
To check our hypothesis, we measured the form factors of \mbox{CS-A} and \mbox{CS-B} microgels after different ultrasonication times using SLS.
SLS was chosen due to the low $q$-range needed for the larger size of the core-shell microgels.
Figures~\ref{fig:SLS_CS}A and \ref{fig:SLS_CS}B show the SLS form factors of \mbox{CS-A} and \mbox{CS-B} microgels, respectively, with increasing $t$ (from bottom to top).
Black lines show the fits using the fuzzy core-shell model~\cite{berndt2005}, which takes into account the different polymer density in the core and shell because of the different crosslinkers.
All fit parameters are shown in Table~S9 (Supporting Information).
Figure~\ref{fig:SLS_CS}C shows the normalized radii of the core-shell microgels, $R/R_0$, obtained from the fits.
Similar to the conventional microgels, $R/R_0$ decrease slower than the normalized hydrodynamic radii, $R_{\textrm{h}}/R_{\textrm{h},0}$, because the dangling chains are removed faster than the more densely crosslinked parts of the microgel network.
Figure~\ref{fig:SLS_CS}D shows the relative polymer density $\phi(r)$ after different $t$, which reflects the structural changes of \mbox{CS-A} microgels.
In the pristine state, the shell crosslinked with BIS is more swollen (has lower polymer density) than the core crosslinked with BAC.
Considering the different resistance of the conventional \mbox{BIS-1} and BAC-1 microgels to ultrasonic degradation (see Figure~\ref{fig:DLS}), one would expect the BIS-crosslinked shell of the \mbox{CS-A} microgels to be removed first followed by degradation of the BAC-crosslinked core.
However, this is not the case, and the core-shell structure persists up to the late stages of degradation (20~min), Figure~\ref{fig:SLS_CS}D (dotted line).
The fuzziness of the \mbox{CS-A} microgels decreases with increasing $t$ similar to the conventional microgels, but only the very periphery of the microgel is affected.
Furthermore, the relative polymer density $\phi(r)$ in the shell (plateau at $r\approx200-250$~nm) remains the same relative to the core.
This means that the shell and the core either degrade simultaneously or not degrade at all.

The \mbox{CS-B} microgels have an inverted structure with respect to the \mbox{CS-A} microgels: the shell crosslinked with BAC is less swollen than the core crosslinked with BIS, Figure~\ref{fig:SLS_CS}E (solid line).
This core-shell structure is also preserved during ultrasonication and only the very periphery is degraded, Figure~\ref{fig:SLS_CS}D (dashed and dotted lines).
So contrary to our initial hypothesis, the more degradable part (core or shell) of a core-shell microgel is not removed selectively by the mechanical force.
Instead, the whole core-shell microgel is affected.
At the same time, the polydispersity also gradually increases with $t$, Figure~\ref{fig:SLS_CS}F.
At the very late stages of degradation ($t>20$~min), Figures~\ref{fig:SLS_CS}A and \ref{fig:SLS_CS}C (upward and left triangles), the scattering profiles become featureless because of high polydispersity.
In this case, the data cannot be fitted unambiguously.

\subsection{Atomic Force Microscopy}

\begin{figure*}[ht!]
    \centering
    \includegraphics[width=\linewidth]{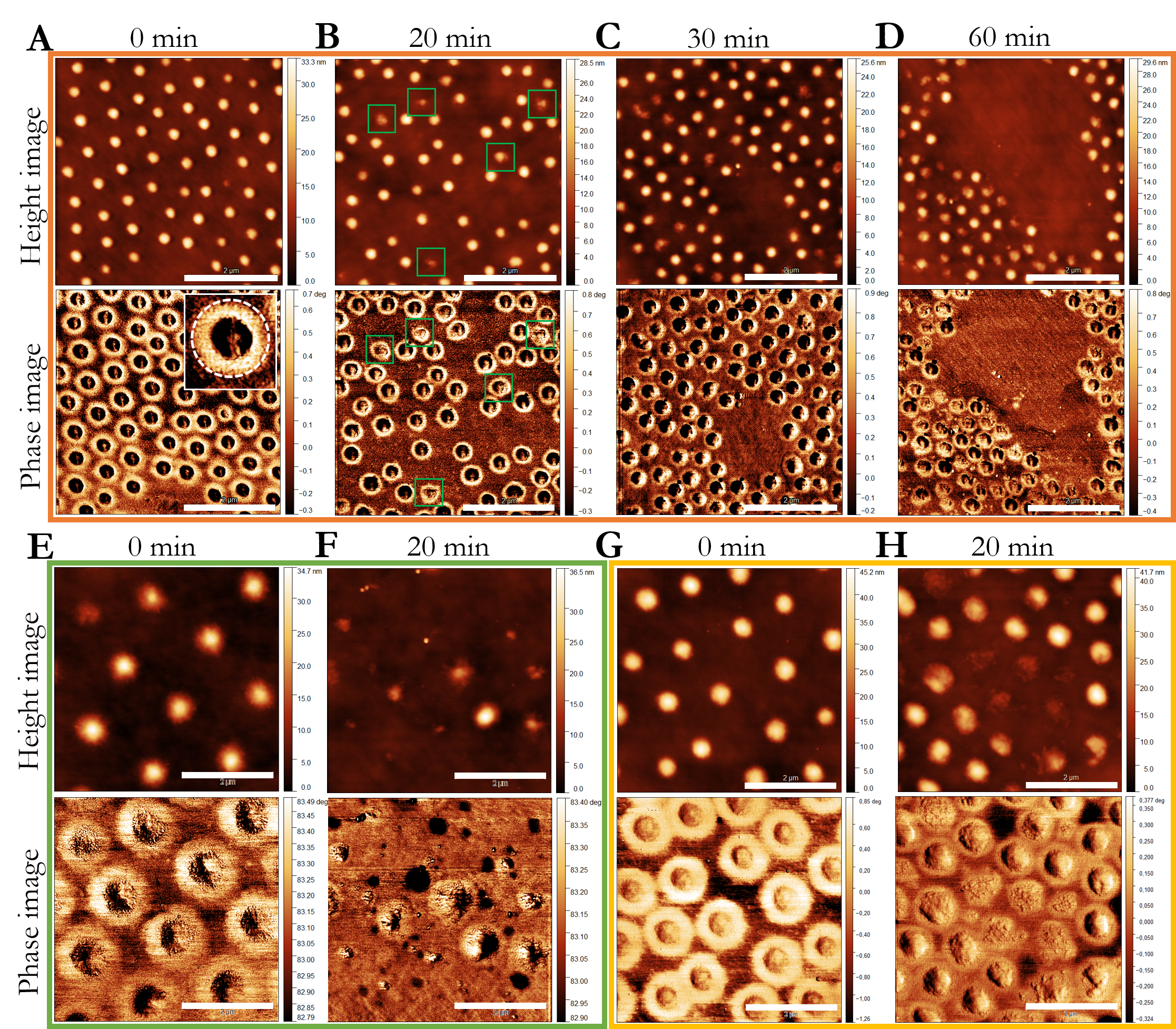}
    \caption{Examples of AFM images of \mbox{BAC-1} microgels (\textbf{A}-\textbf{D}), \mbox{CS-A} microgels (\textbf{E},\textbf{F}), and \mbox{CS-B} microgels (\textbf{G},\textbf{H}) after different ultrasonication times (indicated in the figure). Upper images in each panel are height images, and lower images are phase images. Scale bars are 2~$\mu$m. Inset in panel \textbf{A} (phase image) shows an enlarged microgel and its contact radius $R_{\textrm{cont}}$ (white dashed circle). Green squares in panel \textbf{B} highlight ruptured microgels.}
    \label{fig:AFM_1}
\end{figure*}

As suggested by the SAXS and SLS results, an extensive ultrasonic degradation of microgels eventually leads to non-uniform particles with a broad distribution of sizes or shapes.
Although scattering methods are perfectly suited to determine ensemble average properties, the exact structural changes in each particle leading to the observed non-uniformity are masked by the averaging.
Therefore, we performed Atomic Force Microscopy (AFM) to follow the degradation on the single-particle level.
The microgels were deposited on silicon substrates and imaged in dry state.
First, we focus on \mbox{BAC-1} and \mbox{BIS-5} microgels as examples of microgels that are moderately resistant and highly resistant to ultrasonication, respectively.

Figure~\ref{fig:AFM_1}A-D shows example AFM images for \mbox{BAC-1} microgels after various ultrasonication times: height images (top row) and phase images (bottom row).
The microgels flatten at the interface and show a typical core-corona structure~\cite{bochenek2019}.
In the height images, only the core is observed, while the phase images also resolve the thin corona, which appears as light circles surrounding the core.
The inset of Figure~\ref{fig:AFM_1}A shows a magnified image of a microgel with the core-corona structure.
The white dashed circle shows the radius of the corona that is known as the \textit{contact radius}, $R_{\mathrm{cont}}$.

Only whole microgels and no small fragments or linear chains can be seen in the images.
This indicates that any fragments produced due to microgel degradation are successfully removed during the centrifugation step.
Before ultrasonication, the microgels are monodisperse and radially symmetric, Figure~\ref{fig:AFM_1}A.
However, after 20~min of ultrasonic degradation several irregularly-shaped microgels are observed in the height and phase images, Figure~\ref{fig:AFM_1}B (highlighted with green squares).
At the same time, the rest of the microgels are much less affected by ultrasonication: they remain symmetric and have similar heights as pristine ones.
With increasing ultrasonication time, more of the irregularly-shaped microgels can be seen in the AFM images, Figure~\ref{fig:AFM_1}C (30~min) and Figure~\ref{fig:AFM_1}D (60~min).
Remarkably, even highly degraded microgels retain the corona in the phase images.
\mbox{BIS-5} microgels are affected by ultrasonication in a similar fashion but to a lesser extent, with fewer asymmetric microgels visible in the images, Figure~S9 (Supporting Information).

We propose that the irregular shapes of the microgels are a result of a partial rupture of their network followed by adsorption at the substrate interface.
Strong mechanical forces caused by cavitation can create cracks or weak points in the polymer network.
Upon adsorption, the microgels are further broken by surface forces resulting in highly irregular shapes.
The rupture of microgels is also the most likely reason for the apparent increase of polydispersity after long ultrasonication times, Figure~\ref{fig:SAXS}G.

Next, we probe the asymmetrically crosslinked core-shell microgels using AFM.
Figure~\ref{fig:AFM_1}E and \ref{fig:AFM_1}F show examples AFM images of \mbox{CS-A} microgels before and after 20~min ultrasonication, respectively.
Similarly to the conventional microgels, mechanical degradation caused by the ultrasonication transforms the core-shell microgels into a mixture of irregularly-shaped (ruptured) and relatively intact particles.
In agreement with the SLS data (Figure~\ref{fig:SLS_CS}D and \ref{fig:SLS_CS}E), selective degradation of the shell or core is not observed.
Compared to the conventional \mbox{BAC-1} and \mbox{BIS-5} microgels, the ruptured microgels predominate in the image and also appear much smaller and more irregular in size and shape.
Similarly, the \mbox{CS-B} sample after 20~min ultrasonication shows many irregularly-shaped particles, Figure~\ref{fig:AFM_1}H \textit{vs.} Figure~\ref{fig:AFM_1}G (before ultrasonication).
Compared to the \mbox{CS-A} microgels, the ruptured \mbox{CS-B} microgels assume a more flat ``pancake-like'' morphology with an asymmetric shape.
The difference in the shape of ruptured \mbox{CS-A} and \mbox{CS-B} microgels may be related to preferential rupturing in the shell or core, respectively, and should be the subject of a follow-up study.
The larger fraction of ruptured particles in the AFM images of both \mbox{CS-A} and \mbox{CS-B} microgels compared to the conventional microgels can be explained as follows.
The core-shell microgels are larger than the conventional microgels (Table~\ref{tbl:Rh}), so their ruptured fragments are also larger and, therefore, less likely to be removed during the centrifugation step, see Scheme~\ref{scheme_1}.
The presence of these fragments explains why in DLS the core-shell microgels appear to degrade faster than \mbox{BIS-1} microgels, contrary to the initial expectation (Figure~\ref{fig:DLS}D and \ref{fig:DLS}F).
The intermediate scattering function, $f(q,\tau)$, follows a single-exponential rather than a double-exponential course, which means that the diffusion coefficient of the fragments is not $\gg$ than that of the whole microgels.
Consequently, the apparent $R_{\mathrm{h}}$ values for \mbox{CS-A} and \mbox{CS-B} are the average between the hydrodynamic radii of the whole microgels and the fragments.

\begin{figure*}[htpb!]
    \centering
    \includegraphics[width=\linewidth]{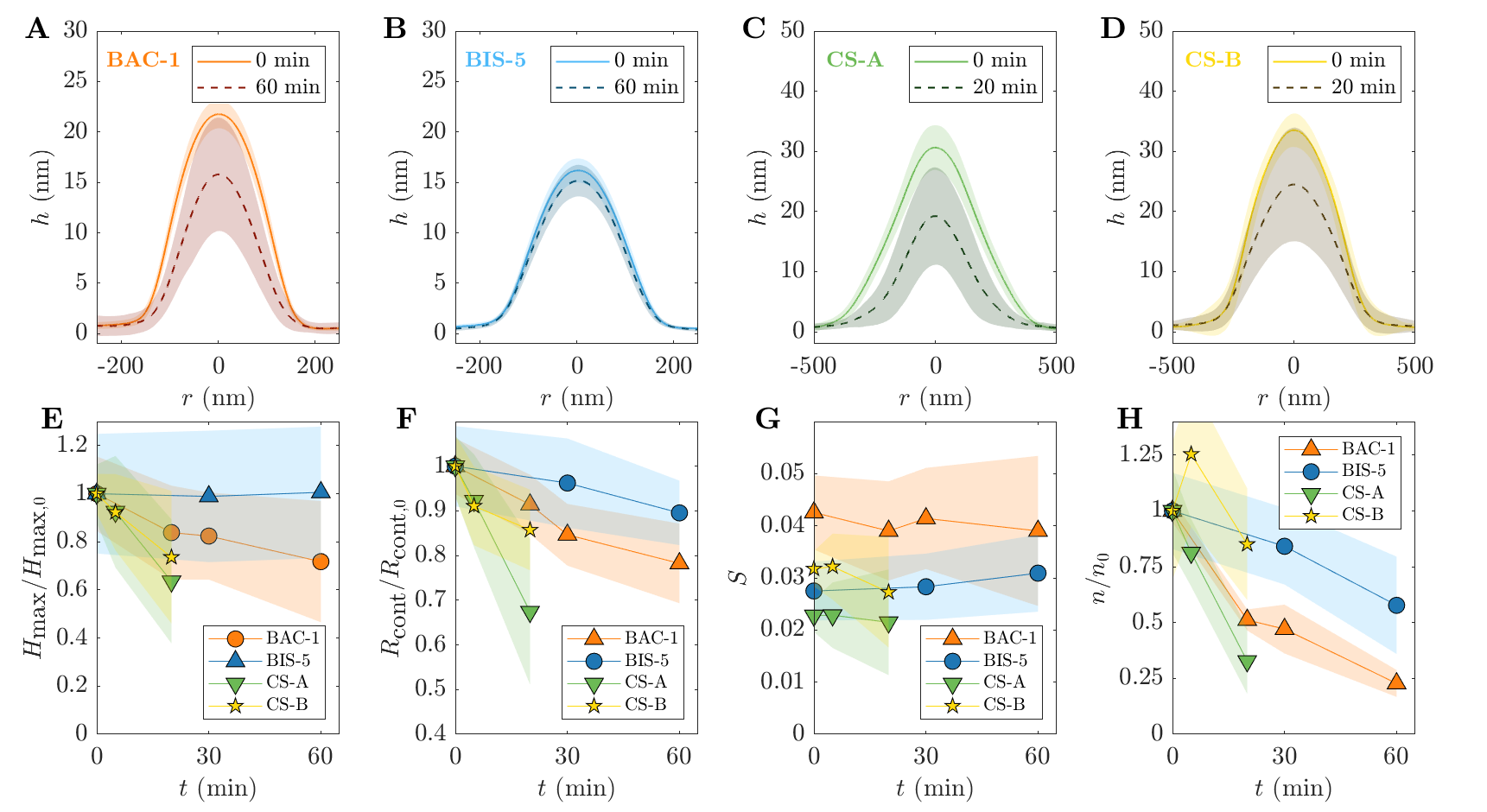}
    \caption{(\textbf{A}-\textbf{D}) Averaged height profiles of \mbox{BAC-1} microgels, \mbox{BIS-5} microgels, \mbox{CS-A} microgels and \mbox{CS-B} microgels (left to right) before and after extensive ultrasonication ($t$ indicated in the legend). (\textbf{E}) Mean maximum height of the microgels, $H_{\textrm{max}}$, normalized by the pristine value, $H_{\textrm{max},0}$, \textit{vs.} ultrasonication time $t$. (\textbf{F}) Mean contact radius of the microgels, $R_{\textrm{cont}}$, normalized by the pristine value, $R_{\textrm{cont},0}$, \textit{vs.} $t$. (\textbf{G}) Mean shape parameter of the adsorbed microgels, $S=H_{\textrm{max}}/2R_{\textrm{cont}}$, \textit{vs.} $t$. (\textbf{H}) Normalized number densities of the microgels in suspension, $n/n_0$, \textit{vs.} $t$. Shaded areas in all panels correspond to standard deviation.}
    \label{fig:AFM_2}
\end{figure*}

Now, we perform a detailed quantitative analysis of the AFM data.
Figures~\ref{fig:AFM_2}A-\ref{fig:AFM_2}D compare averaged height profiles of the microgels before (solid lines) and after extensive ultrasonication (dashed lines) for 60 or 20~min (see legend): \mbox{BAC-1} (A), \mbox{BIS-5} (B), \mbox{CS-A} (C) and \mbox{CS-B} (D).
The maximum height of the profile decreases for the weaker \mbox{BAC-1} microgels, Figure~\ref{fig:AFM_2}A, and remains almost the same for the stronger \mbox{BIS-5} microgels, Figure~\ref{fig:AFM_2}.
At the same time, the standard deviation of both height profiles increases (shaded areas), indicating the increase of apparent polydispersity and appearance of ruptured microgels.
The shape of the profile remains the same (roughly Gaussian) irrespective of $t$.
The core-shell microgels show the same effects but to a larger extent because of their higher susceptibility to mechanical degradation, Figures~\ref{fig:AFM_2}C, \ref{fig:AFM_2}D.
In agreement with the DLS results, \mbox{CS-A} microgels (panel C) show a larger relative decrease in size compared to the \mbox{CS-B} microgels (panel D).
The shape of the profiles, which in this case is different between \mbox{CS-A} and \mbox{CS-B} samples and reflects the different internal structures, remains the same irrespective of $t$.

Figure~\ref{fig:AFM_2}E shows the mean maximum heights of the microgels $H_{\textrm{max}}$ (normalized by the value at 0~min, $H_{\textrm{max},0}$) \textit{vs.} $t$.
Shaded areas correspond to the standard deviation within the sample.
For \mbox{BAC-1} microgels (orange circles), the decrease of $H_{\textrm{max}}$ is statistically significant (one-way ANOVA, $p<0.0001$, Figure~S10A) and is the strongest within the first 20~min of ultrasonication.
Then, the value decreases slower and reaches $H_{\textrm{max}}\approx0.72 H_{\textrm{max},0}$ after 60~min of ultrasonication.
This corresponds to the removal of the periphery of the microgels, including the dangling chains, as also seen from the values of $R_{\textrm{h}}$ and $R_{\textrm{g}}$.
On the other hand, the increase of the standard deviation of $H_{\textrm{max}}$ is likely due to the increasing number of ruptured microgels, which grows significantly at $t=60$~min.
For the \mbox{BIS-5} microgels (blue upward triangles), no significant change of $H_{\textrm{max}}$ is observed (one-way ANOVA, $p=0.852$, Figure~S10B).
Because the standard deviation of $H_{\textrm{max}}$ is already quite high for this sample, the appearance of a few ruptured particles (Figure~S9) cannot be seen from the averaged height data.
The core-shell microgels show a steady decrease of $H_{\textrm{max}}$ at all investigated ultrasonication times, $t\leq20$~min (green downward triangles and yellow stars), which is statistically significant (one-way ANOVA, $p<0.0001$, Figures~S11A and S11D).
The decrease is slightly larger compared to \mbox{BAC-1} microgels.
In this case, it probably results from the combination of the removal of the dangling chains and rupture of microgels, \textit{i.e.} the two contributions cannot be separated in time.

Figure~\ref{fig:AFM_2}F shows the mean radii of microgel corona (see Figure~\ref{fig:AFM_1}A, inset): contact radii $R_{\textrm{cont}}$ \textit{vs.} $t$.
The $R_{\textrm{cont}}$ values are also normalized by the value at 0~min, $R_{\textrm{cont},0}$.
All microgels show a significant decrease of $R_{\textrm{cont}}/R_{\textrm{cont},0}$ with $t$ (one-way ANOVA, $p<0.0001$, Figures~S10B, S10E, S11B and S11E).
The rate of the decrease depends on the susceptibility of a microgel to mechanical degradation: the weaker the microgel the faster is the decrease (\mbox{BIS-5} $<$ \mbox{BAC-1} $<$ \mbox{CS-B} $<$ \mbox{CS-A}).
Since $R_{\textrm{cont}}$ is mostly sensitive to the dangling chains and peripheral parts of the network, this parameter can be a sensitive measure of their removal, complementary to the $\rho$-parameter from DLS/SLS (Figure~\ref{fig:SAXS}C).

Now, the interpretation of the changes in $H_{\mathrm{max}}$ and $R_{\textrm{cont}}$ is not straightforward, because the degradation of microgels can influence their capacity to flatten at interfaces where the AFM measurements are performed~\cite{scotti2022review}.
Therefore, the changes in $H_{\mathrm{max}}$ and $R_{\textrm{cont}}$ could relate to the altered size, internal structure or flattening capacity (softness) of the degraded microgels.
To quantify this capacity, we propose a shape parameter $S$ defined as a quotient of the maximum height $H_{\textrm{max}}$ and the lateral diameter $2R_{\textrm{cont}}$ of a microgel:
\begin{equation}
    S = \frac{H_{\textrm{max}}}{2R_{\textrm{cont}}}.
\end{equation}
The lower the $S$-parameter, the more a microgel flattens at the interface, \textit{i.e.} the softer it is.
Figure~\ref{fig:AFM_2}G shows the mean values of $S$ \textit{vs.} $t$ for the investigated microgels.
In all cases except \mbox{CS-A} microgels, the influence of $t$ on the mean value of $S$ is statistically significant despite the large standard deviations (one-way ANOVA, $p<0.0001$, Figures~S10C, S10F, S11C and S11F).
However, the trends are different for the different microgels.
\mbox{BIS-5} microgels (blue circles) show a gradual increase of $S$ over the whole investigated $t$ interval, whereby the standard deviation of $S$ (shaded area) increases only slightly.
In contrast, \mbox{BAC-1} microgels (orange upward triangles) show first a small decrease of $S$ at $t=20$~min, followed by an equally small increase at $t=30$~min.
At the same time, the standard deviation of $S$ grows significantly with increasing $t$.
\mbox{CS-B} microgels (yellow stars) show a slight decrease of the mean values of $S$ and an increase of its standard deviation similar to \mbox{BAC-1} microgels.
Finally, \mbox{CS-A} microgels (green down triangles) only show an increase of the standard deviation of $S$ with $t$.
To rationalize these non-trivial trends, we note that the removal of the dangling chains and the rupture of microgels are expected to lead to opposite effects on $S$ (increase or decrease, respectively).
Therefore, the contribution of each of the two mechanisms at each $t$ needs to be estimated.

\begin{figure*}[htpb!]
   \centering
   \includegraphics[width=\linewidth]{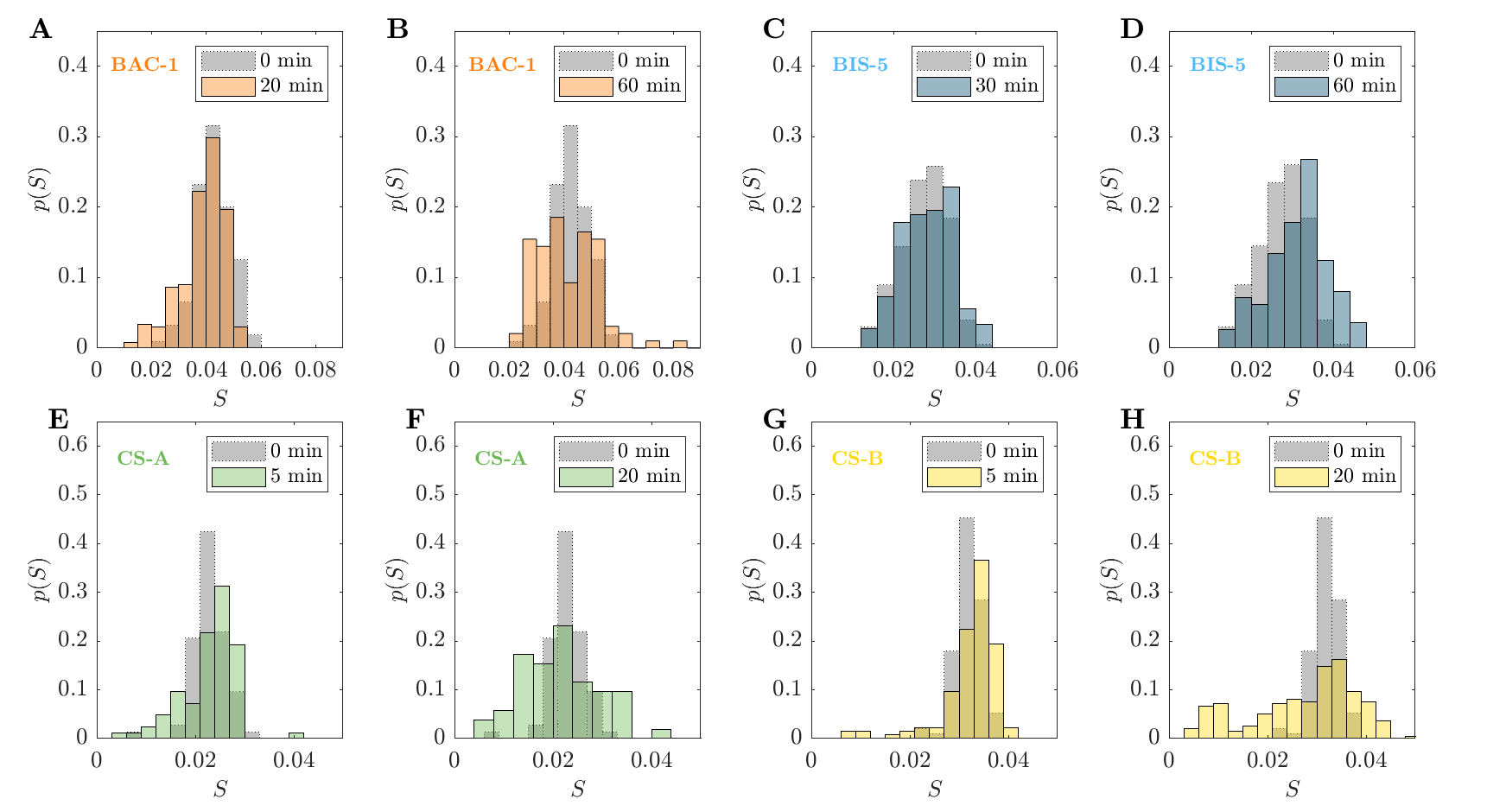}
   \caption{Probability density functions, $p(S)$, of the shape parameter $S$ for \mbox{BAC-1} microgels (\textbf{A},\textbf{B}), \mbox{BIS-5} microgels (\textbf{C},\textbf{D}), \mbox{CS-A} microgels (\textbf{E},\textbf{F}) and \mbox{CS-B} microgels (\textbf{G},\textbf{H}) at different ultrasonication times (indicated in the panels).}
   \label{fig:AFM_S-dist}
\end{figure*}

To achieve this goal, we make use of the particle-tracking algorithms for the analysis of the AFM data (see Experimental section) to calculate the $S$-parameter for each particle individually.
This allows us to obtain the approximate probability density functions (PDFs) of the $S$-parameter, $p(S)$, after different ultrasonication times and detect the changes in the shape of the $S$-parameter distribution.
Figure~\ref{fig:AFM_S-dist} shows the $p(S)$ for different microgels at an early degradation stage (panels A, C, E, G) and a late degradation stage (panels B, D, F, H).
The exact ultrasonication times depend on each microgel's susceptibility to degradation and are indicated in the legend of each panel.
The corresponding PDFs of pristine microgels (0~min) are shown as gray bars with a dotted frame.

For \mbox{BAC-1} microgels at $t=20$~min, Figure~\ref{fig:AFM_S-dist}A, a small shoulder at $S\approx0.02-0.03$ is observed with respect to the $t=0$~min, whereas the maximum of the distribution is not affected.
After extensive degradation for $t=60$~min, Figure~\ref{fig:AFM_S-dist}B, this shoulder develops into the second peak at $S\approx0.03$ resulting in an apparently bimodal distribution of $S$.
At the same time, the first peak shifts to higher $S\approx0.05$.
For \mbox{BIS-5} microgels at the early degradation stage ($t=30$~min), Figure~\ref{fig:AFM_S-dist}C, no significant change in the shape of $p(S)$ can be seen.
After $t=60$~min, Figure~\ref{fig:AFM_S-dist}D, the maximum of the distribution shifts to higher $S$ and an indication of a shoulder at $S\approx0.02$ appears.
\mbox{CS-A} microgels develop a broad bimodal distribution of $S$ already after 5~min of ultrasonication, Figure~\ref{fig:AFM_S-dist}E.
The main peak shifts to higher $S$-values whereas the second peak appears at $S\approx0.018$.
After 20~min ultrasonication, the main peak broadens by developing a shoulder at $S\approx0.03$, while the peak at $S\approx0.018$ becomes more pronounced, Figure~\ref{fig:AFM_S-dist}F.
For \mbox{CS-B} microgels a similar picture is observed: after 5~min the maximum shifts to higher $S$ and a broad shoulder starts to develop at $S\approx0.02$ with an additional small peak at $S\approx0.009$.
After 20~min, the main peak broadens and decreases in intensity, while the shoulder at $S\approx0.02$ and a peak at $S\approx0.009$ grow.

These changes of the PDFs can be assigned to the changes in microgel structure and the extent of its degradation.
The gradual shift of the distribution maximum to higher values of $S$ corresponds to the removal of the dangling chains and outermost parts of the microgel network.
The resulting microgels are effectively harder and can flatten less at the interface, hence the higher values of $S$ without a strong broadening of the distribution.
This effect is seen in all microgels and it is the least affected by the type and amount of crosslinker (susceptibility to degradation).
Conversely, the rupture of microgels corresponds to the formation of a shoulder in $p(S)$ at low values of $S$, which can further develop into the second maximum.
The ruptured microgels are weaker and more affected by the surface forces: adsorption can break them further resulting in very flattened and heterogeneous objects, characterized by low $S$-values with a broad distribution.
The rupture is more pronounced and happens at lower $t$ for microgels that are more susceptible to mechanical degradation, in agreement with our qualitative conclusions from the AFM images.

Finally, we observe that with increasing $t$ larger sample volumes had to be deposited on a substrate to obtain a sufficient number of microgels in the AFM images and have sufficient statistics.
In other words, the number concentration of microgels in the samples decreases with $t$.
Since free polymer chains and fragments are removed by this step and the remaining microgels are re-dispersed in the same amount of water, the exact concentration of our samples is unknown.
However, it can be estimated by simply counting the microgels in the AFM images.
We calculate the number of microgels per unit volume of suspension (number density) $n$ in the samples before and after centrifugation as follows:
\begin{equation}
    n = \frac{\langle N\rangle A_{\textrm{tot}}}{V A_{\textrm{im}}},
\end{equation}
where $\langle N\rangle$ is the average number of microgels per AFM image, $A_{\textrm{tot}}$ is the total area of the interface during deposition (area of the trough), $A_{\textrm{im}}$ is the area of a single AFM image and $V$ is the sample volume deposited on the interface of the trough.

Figure~\ref{fig:AFM_2}H shows the relative number densities $n/n_0$ for the different microgels \textit{vs.} $t$.
For all microgels, even the most stable \mbox{BIS-5}, $n/n_0$ decreases significantly as a result of ultrasonication.
This means that some microgels are destroyed completely and their fragments are removed by centrifugation.
Such a ``catastrophic'' degradation scenario reveals that the distribution of mechanical forces (or strain rates) experienced by individual microgel particles is very broad.
The rate of the $n/n_0$ decrease correlates with the susceptibility to degradation: it is faster for \mbox{BAC-1} compared to \mbox{BIS-5} microgels, $\approx77\%$ and $\approx42\%$ of the particles are destroyed after 60~min, respectively.
This mass loss is significantly higher than from the decrease of the mean size (or molecular weight) of the microgels, which is only $\approx 38\%$ for \mbox{BAC-1} and $\approx 16\%$ for \mbox{BIS-5} microgels after 60~min (see Figure~\ref{fig:DLS}C).
The ``catastrophic'' degradation scenario also explains why for \mbox{BIS-1} microgels no good data could be obtained at $t>10$~min.
Most of the \mbox{BIS-1} microgels, which are the weakest microgels studied here, are completely destroyed after $>10$~min of ultrasonication.

While the \mbox{CS-A} microgels behave as expected, the \mbox{CS-B} microgels show a non-trivial trend of $n/n_0$ with increasing $t$.
They first increase the apparent number density at $t=5$~min and then decrease it at $t=20$~min.
The reason for this can be clearly seen in the AFM images, Figure~\ref{fig:AFM_1}H: many microgel fragments can be seen in the images that are also included in the analysis.
So when a core-shell microgel ruptures into several large fragments, the apparent number density $n$ initially increases.
As the ultrasonication progresses ($t=20$~min), the large fragments are also destroyed leading to a decrease in the apparent value of $n$, as expected.
The fact that such a trend is not observed in \mbox{CS-A} microgels could be related to the different rupturing scenario when the core or shell of a core-shell microgel is weaker.
Indeed, the fragments seen in the AFM images of \mbox{CS-A} microgels after ultrasonication (Figure~\ref{fig:AFM_1}F) are fewer and smaller compared to the fragments of \mbox{CS-B} microgels after a similar $t$ (Figure~\ref{fig:AFM_1}H).
As already mentioned before, the fragments of the conventional microgels (\mbox{BAC-1} and \mbox{BIS-5}) are much smaller and therefore are removed from the suspension by centrifugation.

\section{Discussion}

Our experimental data shows that the degradation of microgels by ultrasound is a complex process.
Unlike linear polymer chains, the susceptibility of a microgel to a mechanical force depends not only on the strength of chemical bonds but also on the network softness, which is closely related to its swelling degree.
Using mechanically-labile crosslinks (like disulfides or diselenides) can indeed make the microgels more susceptible to degradation.
At the same time, these crosslinks can change the network structure and topology, for example the distribution of chain lengths or the number of chain entanglements~\cite{shen2018}, or they can make the polymer more hydrophobic.
A detailed understanding of the network structure is thus needed to predict and rationally implement mechanical responsiveness in microgels.
Here, the future work can build upon the recent progress in designing tough and durable bulk hydrogels~\cite{zhao2021} or elastomers~\cite{ducrot2014} to implement a desired mechanical responsiveness to the network.
Furthermore, many new insights can be gained by comparison of mechanical degradation of microgels with their well-studied chemical degradation~\cite{smith2011,gaulding2012,nishizawa2023,palkar2023}.
In particular, the evolution of size and internal structure, as well molecular weight distribution of fragments could be compared.

Our findings that (i) microgels remain fuzzy particles, \textit{i.e.} have a gradually decaying segment density from the center to the periphery, even after extensive degradation and (ii) that selective degradation of the weaker shell or core is not observed in the core-shell microgels suggest that degradation does not happen on the length scale of individual polymer chains.
Each discrete degradation step, which is related to a single cavitation event, typically involves the breaking of multiple chemical bonds.
In other words, a microgel likely responds to mechanical force as a single elastic body that can deform and break under load.
Therefore, understanding how the stress and strain are distributed across the polymer network in a colloidal (mesoscopic) particle is of paramount importance.
Here, monomer-resolved computer simulations can give unique insights into the fracture mechanism, as for example in the study by~\citeauthor{palkar2023} on chemical degradation of nanogels~\cite{palkar2023} or in the studies of bulk hydrogel fracture~\cite{boots2022,tauber2022}.
Experimentally, a reasonable next step could be to analyze the molecular weight distribution of microgel fragments using, for example, analytical ultracentrifugation~\cite{tauer2009} or fluid flow fractionation methods~\cite{gaulding2012}. 
Designing a microgel with a selectively degradable shell will likely demand that a large difference in the mechanical strength of the core and shell is achieved.
Such that for the average applied force the shell easily breaks but the core is hardly affected. 

\begin{scheme}[htpb!]
    \centering
    \includegraphics[width=0.7\linewidth]{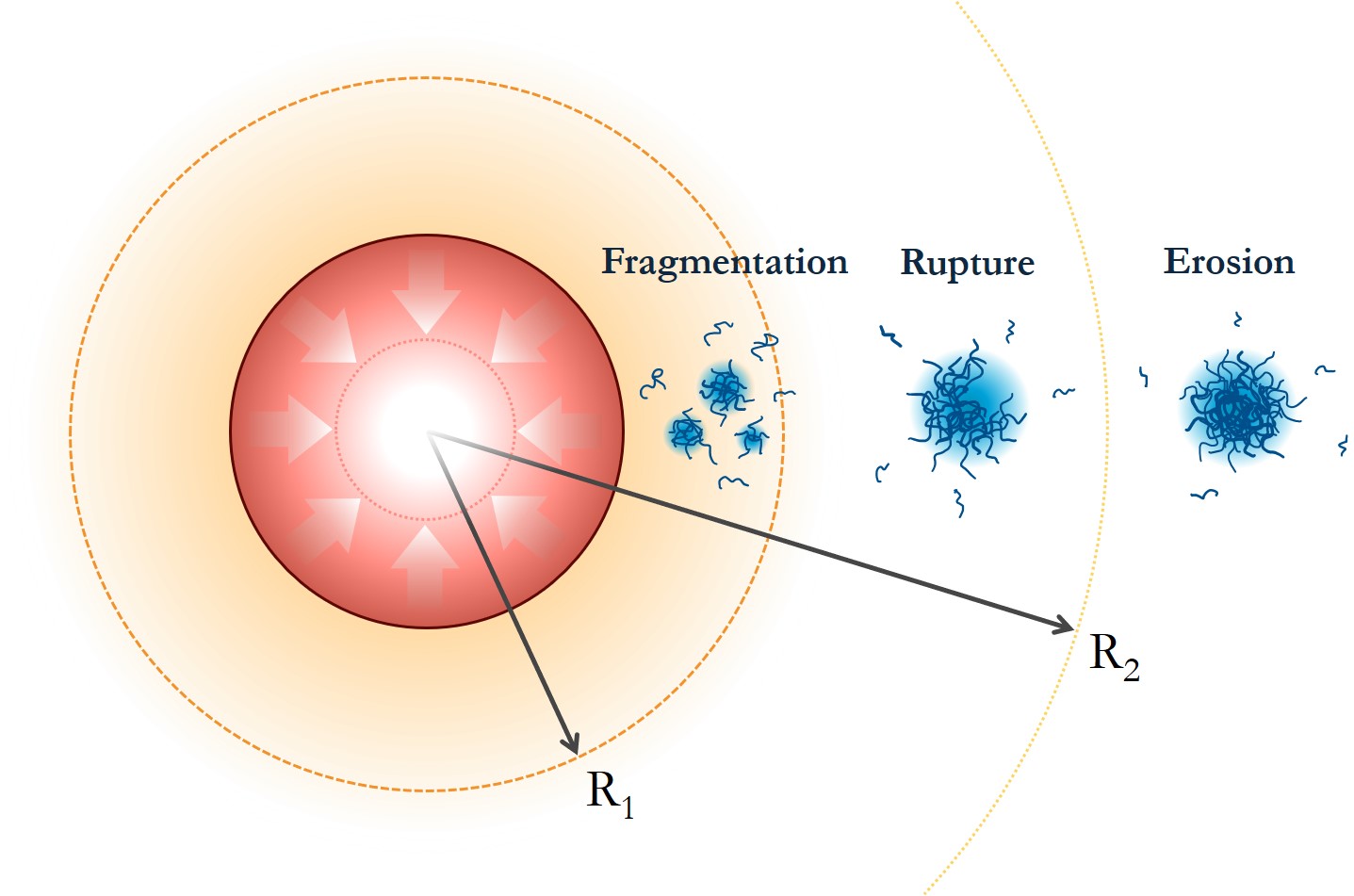}
    \caption{Schematic illustration of the heterogeneous degradation of microgels in low-concentrated dispersion during inertial cavitation. The collapsing cavitation bubble is shown in red. The $R_1$ and $R_2$ are effective cutoff distances for fragmentation and rupture of microgels to occur, respectively.}
    \label{fig:scheme_degrad}
\end{scheme}

Finally, the large heterogeneity in the degradation of microgels by ultrasound is one of the most interesting findings of this study.
The damage experienced by an individual particle over the same ultrasonication time ranges from cleavage of a few dangling chains at the periphery to complete destruction of the particle into fragments.
A similar heterogeneity has been observed and intensely investigated in the context of permealization and lysis of cells by acoustic cavitation~\cite{sundaram2003,guzman2003}.
To explain our results, we propose a qualitative framework similar to the one used by \citeauthor{sundaram2003} and \citeauthor{guzman2003} in the cell experiments.
First, we consider that the main phenomena responsible for exerting mechanical forces are the wall motion during the expansion-collapse cycle of the cavitation bubble and the shock waves emitted after its collapse.
The impact of both effects decay with increasing distance from the bubble center $r$: the pressure amplitude of the shock wave decreases as $1/r$ , while the strain due to the bubble wall motion decreases as $1/r^3$~\cite{sundaram2003}.
Now, for a single cavitation event, we can imagine two (qualitative) cutoff distances from the center of the bubble, $R_1$ and $R_2$ , Scheme~\ref{fig:scheme_degrad}, such that: (i) a microgel located at $r<R_1$ is completely destroyed into fragments, (ii) a microgel located at $R_1<r<R_2$ is ruptured or forms a crack in its polymer network and (iii) a microgel located at $r>R_2$ can only undergo a slow erosion by losing its dangling chains.
The parameters $R_1$ and $R_2$ are expected to depend on the nature of microgels: the weaker the microgels, the higher $R_1$ and $R_2$ values should be for a given ultrasound intensity and concentration of air bubbles.
The kinetic model for this approach proposed by~\citeauthor{sundaram2003} predicts the number of particles to decrease exponentially with $t$, which is approximately what we observe in Figure~\ref{fig:AFM_2}H.
The deviations from the exponential course are likely due to inclusion of microgel fragments in the analysis and a large experimental error.

The ``erosion'' mechanism is most likely responsible for the removal of the dangling chains and microgel periphery, observed by DLS, SLS, AFM ($R_{\textrm{cont}}$) and discussed in several previous studies~\cite{izak-nau2020,kharandiuk2022,he2023}.
This mechanism affects the majority of microgels because it requires the scission of only one covalent bond at a time.
However, the ``erosion'' mechanism also accounts for only a moderate mass loss, see Figure~\ref{fig:DLS}C.

The ``rupture'' mechanism is manifested in the results of form factor analysis by the decrease of $R$ and increase of $p$, as well as in the AFM images by the irregularly-shaped particles and in the $p(S)$ distributions as the second maximum at low $S$.
This mechanism requires the scission of multiple bonds at once.
Therefore, it is less likely to occur than erosion and affects fewer microgels.
It is also possible that single broken bonds accumulate in a microgel over the course of ultrasonication and, after some time, result in a rupture event.
This scenario would explain why ruptured particles in AFM images and significant changes to the microgel form factors are seen only during the latter stages of degradation.

Lastly, the ``fragmentation'' mechanism is responsible for the decrease of number concentration of microgels, $n$, and accounts for the largest polymer mass loss among the three mechanisms.
Measurements of $n$ are therefore required in the future studies, for example, using the method proposed by~\citeauthor{hildebrandt2022}~\cite{hildebrandt2022}
This mechanism requires the most bond scissions at a time and should therefore be very rare, \textit{e.g.} only in the immediate vicinity of a collapsing cavitation bubble.
However, it is possible that fragmentation also occurs \textit{via} several consecutive rupture events, which might increase its probability.
In order to verify if this is indeed the case, a detailed study of the molecular weight distribution is required for both degraded microgels and their fragments.

The above classification offers a simplified picture based on our interpretation of the AFM and scattering data obtained after different sonication times for microgels purified using ultracentrifugation.
For a more comprehensive understanding, the small fragments removed by the ultracentrifugation step should be carefully analyzed and included in the discussion.
However, we believe that the proposed picture provides a useful framework to explain different degradation scenarios and design new experiments on mechanically-responsive colloids.
Further studies could, for example, develop classification algorithms to extract the fractions of intact and degraded microgels from AFM data.
Alternatively, one could think of using synthetic microbubbles under stable cavitation conditions to achieve a more homogeneous and controllable strain field~\cite{guzman2003}.
The mechanical force exerted due to bubble wall motion and microstreaming~\cite{nyborg1982} under such conditions might be sufficient to break the weakest bonds: strain rates of $\approx10^3$~s$^{-1}$ have been reported at a bubble oscillation amplitude as low as $5\%$ of the bubble radius \cite{marmottant2003}.

\section{Conclusion}

In this study, we use a combination of several scattering techniques and atomic force microscopy to systematically investigate the mechanism of microgel degradation by high-intensity ultrasound and the role of mechanoresponsive disulfide crosslinks in this process.
We show that the susceptibility of a microgel to mechanical degradation results from an interplay between the presence of mechanoresponsive bonds and the swelling degree of the polymer network.
Furthermore, we find that the core-shell microgels with different crosslinks in the core and shell cannot be degraded selectively in their weaker part.
Instead, the core-shell structure is affected as a whole, and its susceptibility to degradation is close to that of its weaker part.
The mechanism of degradation reflects the large heterogeneity of mechanical forces imposed by the cavitation bubbles on the microgels.
The majority of microgels are located far away from the cavitation bubbles and, therefore, experience a relatively weak force.
This force leads to a slow erosion of their periphery – dangling chains or weakly crosslinked shell – which can be detected using both light scattering and AFM.
In contrast, a small fraction of microgels that are close to a cavitation bubble experience a much stronger force, leading to their rupture or complete disintegration into small fragments.
Partial rupture results in microgels with a smaller size and a higher apparent size polydispersity, as seen by the form factor analysis.
Furthermore, the ruptured microgels deform stronger upon adsorption at (solid) interfaces than pristine microgels and assume asymmetrical shapes, easily detectable by AFM image analysis.
Disintegration of microgels into small fragments is responsible for a significantly higher mass loss compared to both the erosion of the microgel periphery and the partial rupture of their polymer network.
This mechanism remains largely unexplored in the literature because a measurement of particle number concentration is needed to confirm it.
Overall, our findings highlight the high complexity in mechanical degradation of polymeric microgels, which bears similarities with both the degradation of linear polymers and failure of bulk hydrogels.
Our observations can inspire further studies of the extreme mechanics of mesoscopic polymer particles like microgels.
Such studies will provide a solid background for the rational design of ultrasound-responsive drug delivery systems and help to address the problems related to formation and spreading of microplastics.

\section{Author Contribution Statement}

Alexander V. Petrunin: conceptualization, investigation (DLS, SLS, SAXS, AFM, ultracentrifugation), formal analysis, data curation, writing - original draft; Susanne Braun: conceptualization, investigation (synthesis, DLS, reaction calorimetry, Raman spectroscopy, ultrasonication), formal analysis, writing - original draft; Felix J. Byn: investigation (DLS, SLS, AFM), formal analysis; Indr\'{e} Milvydait\'{e}: investigation (synthesis); Timon Kratzenberg: software (AFM analysis script); Pablo Mota-Santiago: investigation (SAXS); Andrea Scotti: investigation (SAXS), supervision, writing - review \& editing; Andrij Pich: conceptualization, funding acquisition, project administration, supervision, writing - review \& editing; Walter Richtering: conceptualization, funding acquisition, project administration, supervision, writing - review \& editing.

\section{Data Availability Statement}

All data used in the manuscript are available from the authors upon reasonable request.

\begin{acknowledgement}
The authors thank the Deutsche Forschungsgemeinschaft for financial support within the SFB 985 ``Functional microgels and microgel systems''.
Andrea Scotti acknowledges financial support from the Knut and Alice Wallenberg Foundation (Wallenberg Academy Fellows) and from the Swedish Research Council (Research Grant 2024-04178).
The SAXS measurements were performed on the CoSAXS beamline at the MAX IV laboratory (Lund, Sweden) under the proposal 20220526.
The Research conducted at MAX IV, a Swedish national user facility, is supported by the Swedish Research council under contract 2018-07152, the Swedish Governmental Agency for Innovation Systems under contract 2018-04969, and Formas under contract 2019-02496.
\end{acknowledgement}

\begin{suppinfo}

Amounts of chemicals used for the synthesis of microgels; Raman spectroscopy data and the calibration curve; In-line reaction calorimetry data; Discussion of the difference in swelling degree between BIS- and BAC-crosslinked microgels; Additional DLS, SLS, SAXS data and fit parameters; Additional AFM data.

\end{suppinfo}
\newpage
\bibliography{refs}

\end{document}



\newpage
\section{S1. Amounts of Chemicals and Other Details to the Syntheses of Microgels}

\begin{table}[ht!]
    \centering
    \begin{tabular}{cc}
    \hline
    Microgel & Sample code in the lab journal \\
    \hline
    BIS-1 & SB-C6  \\
    BAC-1 & SB-1BAC \\
    BIS-5 & SB-5BIS  \\
    BAC-5 & SB-5BAC \\ 
    CS-A & IM-12 \\
    CS-B & IM-18 \\
    \hline
    \end{tabular}
    \caption{Internal sample codes (lab journal) of the microgels used in the main manuscript.}
    \label{tbl:sample_codes}
\end{table}

\begin{table}[ht!]
  \caption{Exact amounts of chemicals used for the syntheses of the conventional microgels used in the main manuscript.}
  \label{tbl:Syn_main}
  \resizebox{\textwidth}{!}{%
  \begin{tabular}{cccccccccc}
    \hline
    Name & Crosslinker & Reaction & $m$(NIPAm) & $m$(BIS)  & $m$(BAC) & $m$(SDS) & $m$(TEMED) & $m$(APS) & $V$(H$_2$O) \\
         &  content & vessel & [g] &  [g] &  [g] & [g] & [g] & [g] & [mL] \\
    \hline
    BIS-1 & 1~mol\% BIS & flask &  &  & - &  & &  &  \\
    BIS-5 & 5~mol\% BIS & calorimeter & &  & - &  & &  &  \\
    BAC-1 & 1~mol\% BAC & calorimeter & 4.0000 & - & 0.0921 & 0.0417 & 0.0411 & 0.0810 & 300.79 \\
    BAC-5 & 5~mol\% BAC & calorimeter & 4.0000 & - & 0.4701 & 0.0425 & 0.0411 & 0.0805 & 300.08\\
    \hline
  \end{tabular}}
\end{table}

\begin{table}[ht!]
  \caption{Exact amounts of chemicals used for the syntheses of conventional microgels used as cores for the core-shell microgels.}
  \label{tbl:Syn_C}
  \resizebox{\textwidth}{!}{%
  \begin{tabular}{ccccccccc}
    \hline
   Name & Crosslinker & $m$(NIPAm) & $m$(BIS)  & $m$(BAC) & $m$(SDS) & $V$(TEMED) & $m$(APS) & $V$(H$_2$O) \\
         &  content & [g] &  [g] &  [g] & [g] & [µL] & [g] & [mL] \\
    \hline
    CS-A core & 1~mol\% BAC & 2.001 & - & 0.048 & 0.021 & 26.5 & 0.041 & 150 \\
    CS-B core & 1~mol\% BIS & 1.087 & 0.015 & - & 0.017 & 14.1 & 0.022 & 80 \\
    
    \hline
  \end{tabular}}
\end{table}

\begin{table}[ht!]
  \caption{Characteristics of the conventional microgels used as cores for the core-shell microgels.}
  \label{tbl:Size_C}
  \resizebox{\textwidth}{!}{%
  \begin{tabular}{cccccccc}
    \hline
    Name & Crosslinker & $c$(microgel)  & yield  & $R_{\textrm{h}}(20\;^{\circ}\textrm{C})$ & PDI & $R_{\textrm{h}}(60\;^{\circ}\textrm{C})$& PDI \\
      & content &  [mg$\cdot$ mL$^{-1}$] &  [\%] &  [nm] & [a.u.] & [nm] & [a.u.]\\
    \hline
    CS-A core & 1~mol\% BAC & 12.7 & 90.3 & 124 $\pm$ 1.46 & 0.0136 $\pm$ 0.0074 & 68 $\pm$ 0.74 & 0.0152 $\pm$ 0.0106 \\
    CS-B core & 1~mol\% BIS & 10.7 & 75.4 & 134 $\pm$ 2.77 & 0.0103 $\pm$ 0.0080 & 53 $\pm$ 0.42 & 0.0177 $\pm$ 0.0118 \\
    \hline
  \end{tabular}}
\end{table}

\begin{table}[ht!]
  \caption{Exact amounts of chemicals used for the syntheses of the shell of the core-shell microgels.}
  \label{tbl:Syn_CS}
  \resizebox{\textwidth}{!}{%
  \begin{tabular}{ccccccccc}
    \hline
    Name & Crosslinker & $V$(core solution) & $m$(NIPAm) & $m$(BIS) & $m$(BAC) & $V$(TEMED) & $m$(APS) & $V$(H$_2$O) \\
        & content & [mL] & [g] &  [g] &  [g] & [µL] & [g] & [mL] \\
    \hline
    CS-A & 1~mol\% BAC/1~mol\% BIS &  & 1.201 & 0.016 & - & 15.9 & 0.024& 60 \\
    CS-B & 1~mol\% BIS/1~mol\% BAC &  & 1.201 & - & 0.028 & 15.9 & 0.025 & 60 \\
    \hline
  \end{tabular}}
\end{table}




\newpage
\section{S2. Determination of BAC Content in the Microgels Using Raman Spectroscopy}

\begin{figure}[ht!]
    \centering
    \includegraphics[width=\linewidth]{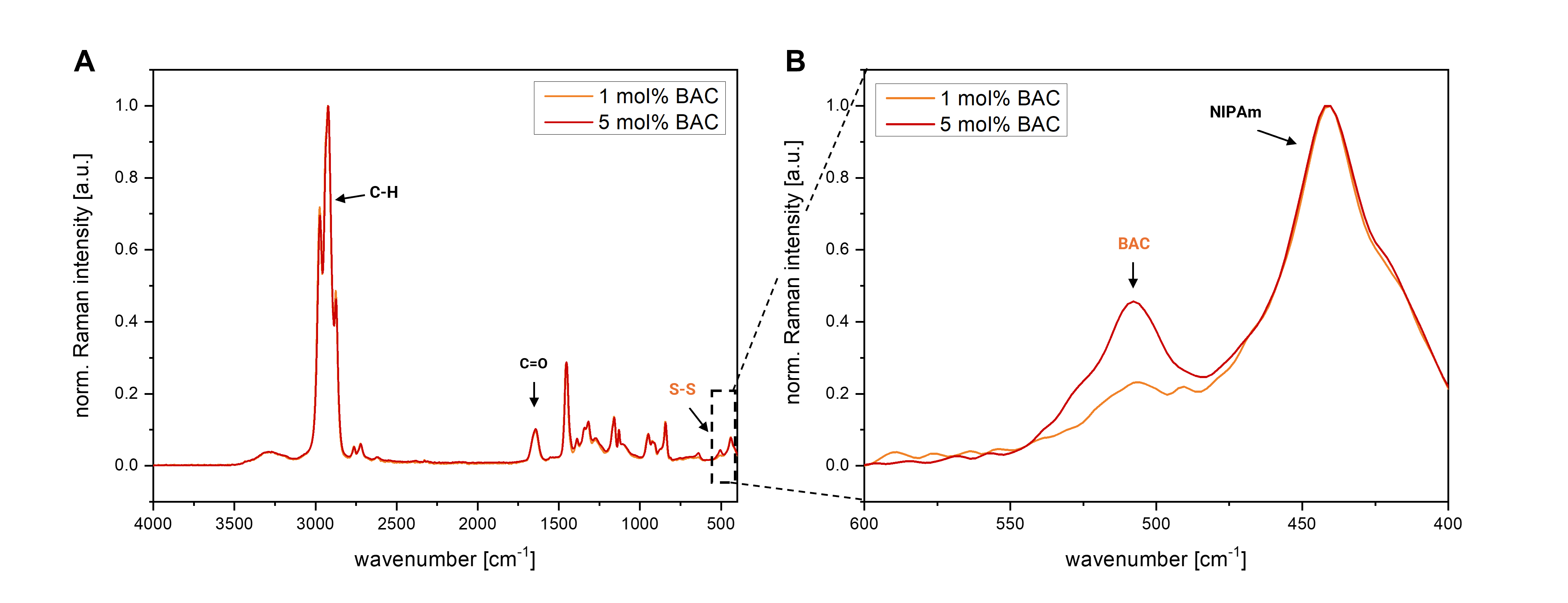}
    \caption{Fourier Transform Raman spectra of BAC-crosslinked pNIPAm microgels with 1 and 5\;mol\% of theoretical crosslinker content. (\textbf{A}) Total Raman spectrum, and (\textbf{B}) enlarged view on the characteristic disulfide-bond region.}
    \label{fig:Raman-samples}
\end{figure}

\begin{figure}[ht!]
    \centering
    \includegraphics[width=0.65\linewidth]{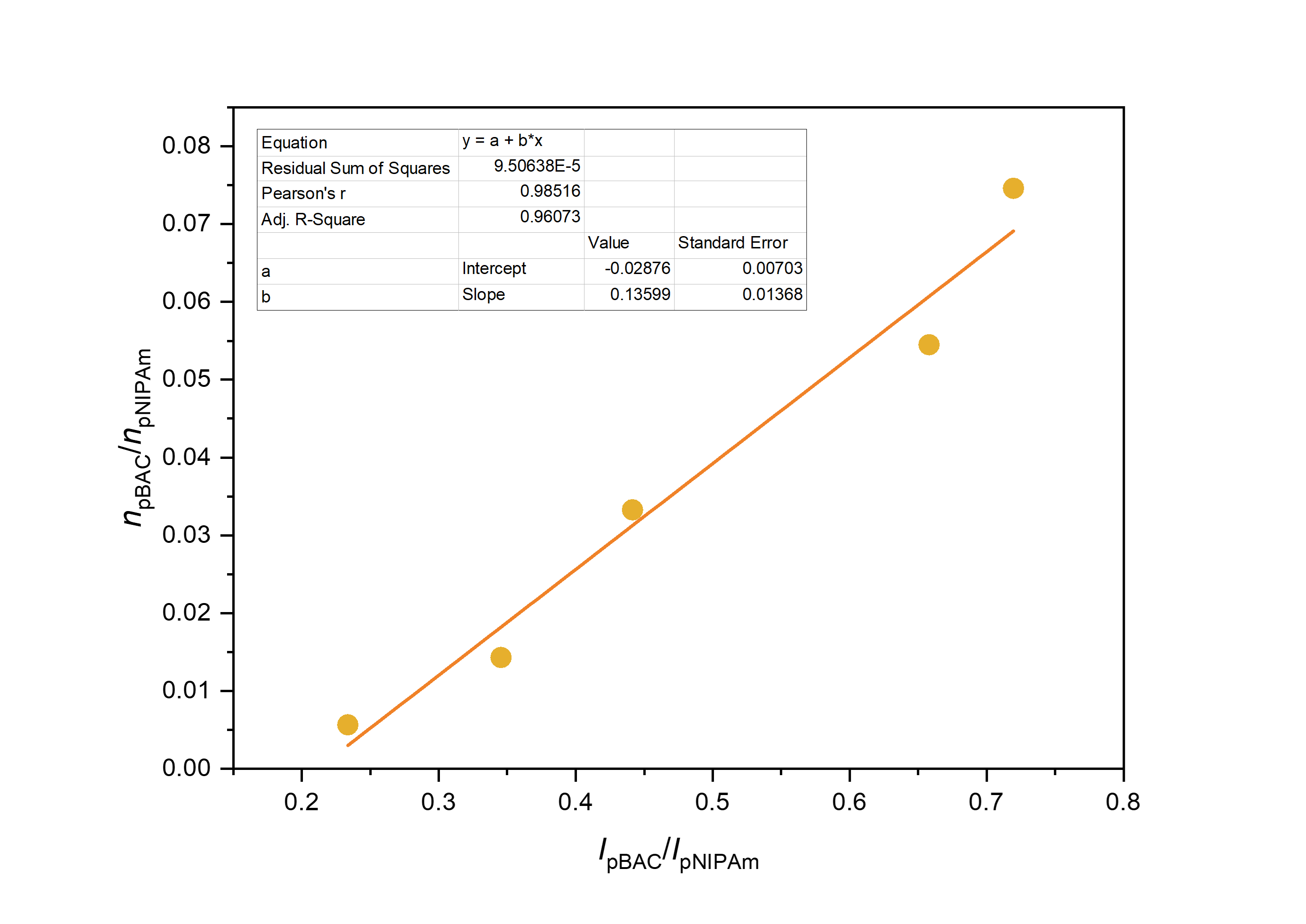}
    \caption{Raman calibration curve used to quantify the BAC content incorporated into the microgel network and obtained by mixing of homopolymers of pNIPAm and pBAC.}
    \label{fig:Raman-calibration}
\end{figure}

In order to quantify how much disulfide bonds remain in the BAC-crosslinked microgels, we performed Raman spectroscopy measurements.
The amount of S-S groups was calculated from their characteristic peak at $\approx 500$ cm$^{-1}$, Figure~\ref{fig:Raman-samples}, using a calibration curve obtained by mixing known amounts of homopolymers (pNIPAm and pBAC), Figure~\ref{fig:Raman-calibration}.
For the microgels with nominal amounts of BAC crosslinker equal to 1, 3 and 5 mol\%, we found the actual BAC contents of 0.3, 1.8 and 3.2 mol\%, respectively.

\newpage
\section{S3. Monitoring the Kinetics of the Microgel Syntheses Using In-Line Reaction Calorimetry}

\begin{figure}[ht!]
    \centering
    \includegraphics[width=\linewidth]{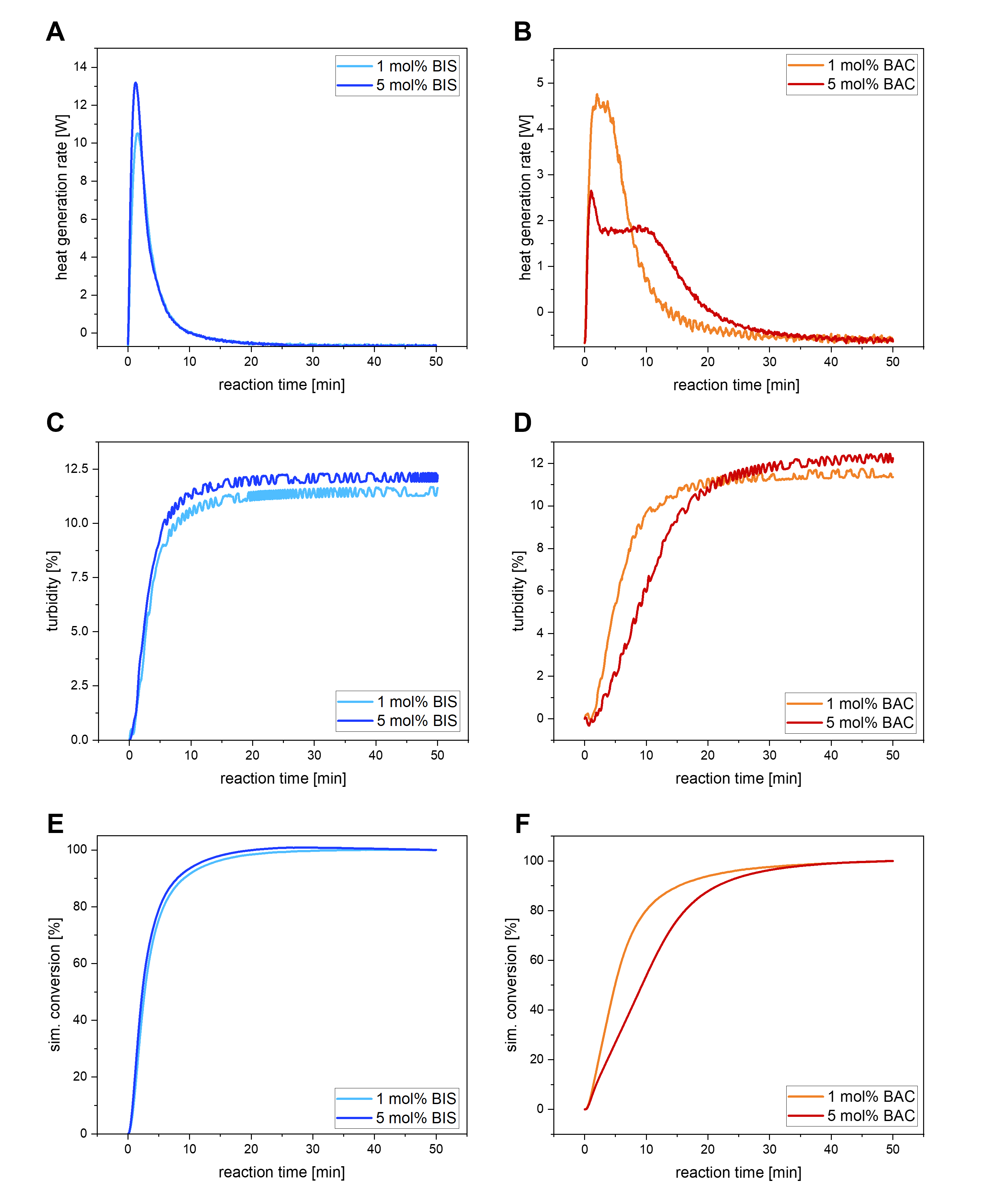}
    \caption{Comparison of the microgel synthesis of pNIPAm/BIS and pNIPAm/BAC microgels initiated with the redox initiation system of APS/TEMED with various crosslinker concentrations of 1 and 5\;mol\% using in-line reaction calorimetry. Heat generation rate of (\textbf{A}) BIS-crosslinked, and \textbf{B}) BAC-crosslinked; turbidity of (\textbf{C}) BIS-crosslinked, and (\textbf{D}) BAC-crosslinked; related simulated conversion of (\textbf{E}) BIS-crosslinked, and (\textbf{F}) BAC-crosslinked microgels.}
    \label{fig:reaction-calorimetry}
\end{figure}

To better understand the difference between BIS- and BAC-crosslinked microgels, their polymerization kinetics were followed using in-line reaction calorimetry.
Figures~\ref{fig:reaction-calorimetry}A and~\ref{fig:reaction-calorimetry}B show the heat generation rate after the start of polymerization as a function of time for BIS-crosslinked microgels (A) and BAC-crosslinked microgels (B).
BIS-crosslinked microgels show a single sharp peak, which shifts to lower times with increasing concentration of BIS (1~mol\% \textit{vs.} 5~mol\%), Figure~\ref{fig:reaction-calorimetry}A.
This peak corresponds to copolymerization of BIS and NIPAm, and the shift is due to a faster polymerization rate of BIS with respect to NIPAm.
In contrast, BAC-crosslinked microgels show a much broader peak, which splits into two overlapping peaks with increasing the BAC concentration from 1~mol\% to 5~mol\%, Figure~\ref{fig:reaction-calorimetry}B.
Since the polymerization rate of BAC is higher compared to NIPAm, the first peak can be assigned to preferential homopolymerization of BAC, whereas the broad second peak likely corresponds to copolymerization of BAC and NIPAm.
Significant broadening of the heat generation rate peak in BAC-crosslinked microgels compared to the BIS crosslinked microgels can be assigned to side-reactions of the S-S group.
Such side-reactions can lead to transient formation of (more stable) S$\cdot$ radicals, which might have a slower chain propagation rate or earlier termination of a growing chain.
Indeed, the content of S-S groups in the final microgels is systematically lower than the nominal BAC amounts, as shown by the Raman spectroscopy results (Figures~\ref{fig:Raman-samples} and~\ref{fig:Raman-calibration}).

The completion of the polymerization reaction was verified by following the turbidity of the solutions, Figures~\ref{fig:reaction-calorimetry}C and~\ref{fig:reaction-calorimetry}D, as well as by simulating the conversion using the iControl RC1e 5.3 software, Figures~\ref{fig:reaction-calorimetry}E and~\ref{fig:reaction-calorimetry}F.
For all microgels, the polymerization was complete within 60~min after adding the initiator, \textit{i.e.} no change of either turbidity or simulated conversion was observed.



\newpage
\section{S4. Influence of the BAC Crosslinker on the Swelling Degree}
\label{sec:crosslinker_swelling}

The swelling degrees for BAC-1 and BAC-5 microgels (Table~1, Main Manuscript) are significantly lower than what is expected for pNIPAm microgels with similar amounts of crosslinker~\cite{lopez2017}.
This gives further evidence that part of the BAC crosslinker was consumed in side-reactions, resulting in formation of permanent crosslinks (C-S bonds)~\cite{gaulding2012} and a higher actual number of crosslinks than expected.

However, the difference in swelling between BIS-1 and BAC-1 microgels cannot be explained only by the increase of the number of crosslinks due to S-S bond dissociation.
If we consider that every dissociated S-S crosslink produces two C-S crosslinks and that 0.3~mol\% of the S-S bonds remain in the microgels (based on Raman spectroscopy data), then the real number of crosslinks should be:
\begin{equation}
    x=(1~\textrm{mol}\%-0.3~\textrm{mol}\%)\cdot2 + 0.3~\textrm{mol}\%=1.7~\textrm{mol}\%.
\end{equation}
Although the real number of crosslinks is then 70\% higher than their nominal number, it is still relatively low and is not sufficient to explain the low swelling degree.
Indeed, the swelling degree of BAC-1 microgels, $\alpha_{\textrm{BAC-1}}=1.92\pm0.02$, is lower than that of the BIS-5 microgels, $\alpha_{\textrm{BIS-5}}=2.11\pm0.04$, which contain 5~mol\% of crosslinks.
Other reasons, such as the high hydrophobicity of BAC and a different distribution of crosslinks (network topology), must also contribute to the low swelling degree of BAC-crosslinked microgels.

\newpage
\section{S5. Additional Scattering Data and Fit Parameters (DLS, SLS, SAXS)}

\begin{figure}[h!]
    \centering
    \includegraphics[width=\linewidth]{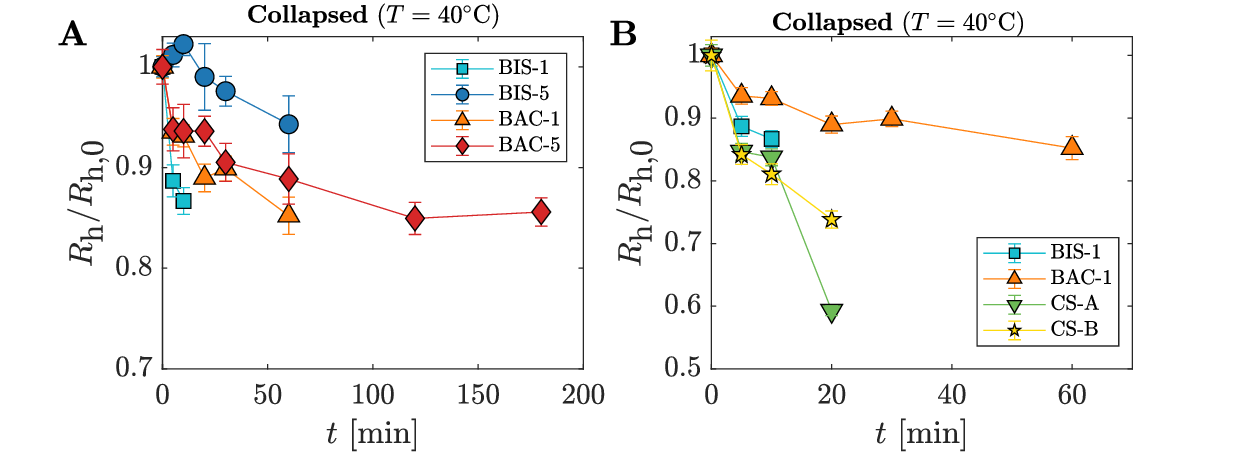}
    \caption{Normalized hydrodynamic radii of the microgels in collapsed state $R_{\textrm{h}}/R_{\textrm{h},0}(T=40^{\circ}\textrm{C})$ for: (A) simple microgels and (B) hybrid core-shell microgels.}
    \label{fig:DLS_SI}
\end{figure}

\begin{figure}[h!]
    \centering
    \includegraphics[width=\linewidth]{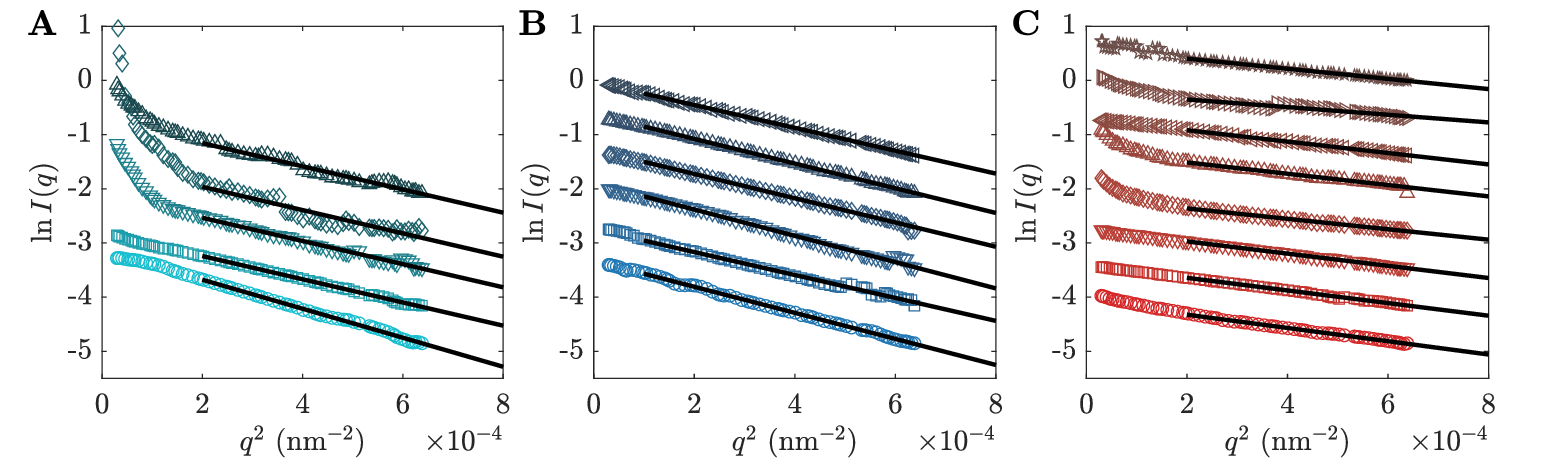}
    \caption{Guinier plots from SLS after different ultrasonication times $t$ for: (A) BIS-1 microgels after 0~min, 5~min, 10~min, 30~min, 60~min ultrasonication (from bottom to top); (B) BIS-5 microgels after 0~min, 5~min, 10~min, 20~min, 30~min, 60~min ultrasonication (from bottom to top); (C) BAC-5 microgels after 0~min, 5~min, 10~min, 20~min, 30~min, 60~min, 120~min, 180~min ultrasonication (from bottom to top). The curves are shifted along the $y$-axis for clarity. Black solid lines correspond to linear fits.}
    \label{fig:SLS_SI}
\end{figure}

\newpage
\begin{figure}[h!]
    \centering
    \includegraphics[width=0.5\linewidth]{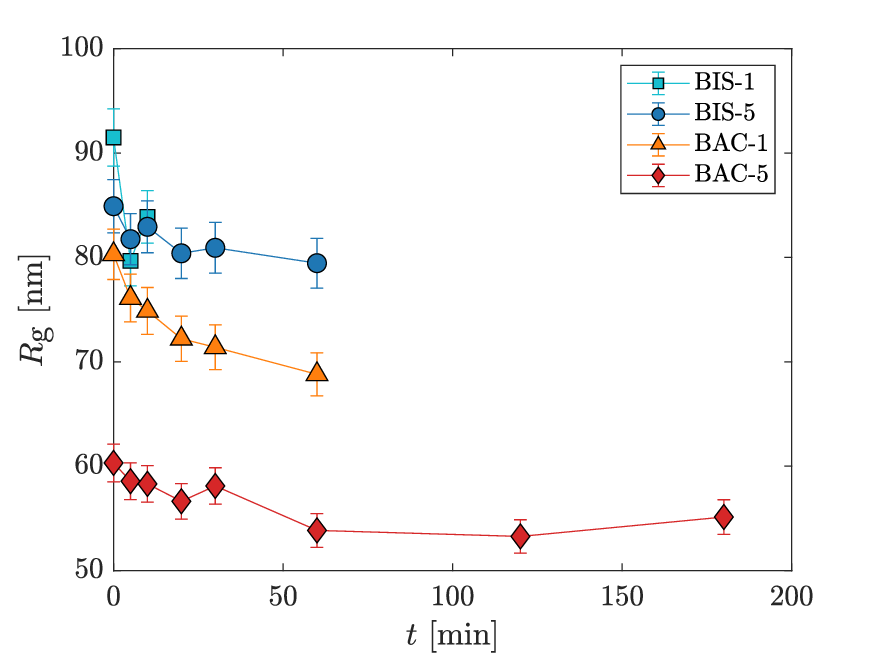}
    \caption{Radii of gyration $R_{\textrm{g}}$ for the simple microgels \textit{vs.} ultrasonication time $t$.}
    \label{fig:Rg_SI}
\end{figure}

\begin{figure}[h!]
    \centering
    \includegraphics[width=\linewidth]{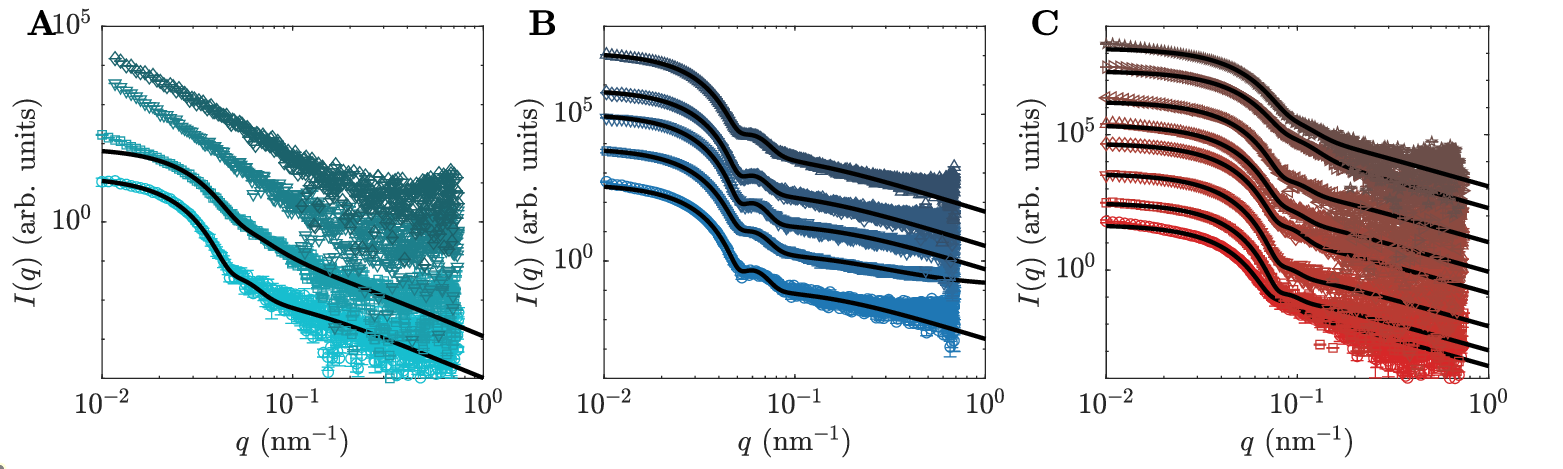}
    \caption{SAXS intensties $I(q)$ \textit{vs.} scattering vector $q$ after different ultrasonication times $t$ for: (A) BIS-1 microgels after 0~min, 5~min, 10~min, 20~min ultrasonication (from bottom to top); (B) BIS-5 microgels after 0~min, 5~min, 10~min, 20~min, 30~min ultrasonication (from bottom to top); (C) BAC-5 microgels after 0~min, 5~min, 10~min, 20~min, 30~min, 60~min, 120~min, 180~min ultrasonication (from bottom to top). The curves are shifted along the $y$-axis for clarity. Black solid lines correspond to fits using the fuzzy sphere model.}
    \label{fig:SAXS_SI}
\end{figure}

\newpage
\begin{table}[htpb!]
  \caption{Fit parameters of the fuzzy sphere form factor model obtained for the simple microgels (SAXS data) after different ultrasonication times $t$.}
  \label{tbl:fit_param}
  \begin{tabular}{ccccccc}
    \hline
    Microgel & $t$ & $R$ [nm] & $R_{\textrm{c}}$ [nm] & $2\sigma$ [nm] & $p$ [\%] & $\xi$ [nm]\\
    \hline
    \multirow{2}{*}{BIS-1}  & $0$ & $130\pm6$ & $42\pm2$ & $87\pm4$ & $17\pm3$ & $12\pm2$ \\
                            & $5$ & $83\pm4$ & $50\pm2$ & $33\pm2$ & $30\pm5$ & $15\pm2$ \\
    \hline
    \multirow{6}{*}{BAC-1}  & $0$ & $121\pm6$ & $33\pm2$ & $88\pm4$ & $10\pm1$ & $15\pm2$ \\
                            & $5$ & $113\pm6$ & $40\pm2$ & $72\pm4$ & $12\pm2$ & $12\pm2$ \\
                            & $10$ & $119\pm6$ & $31\pm2$ & $88\pm4$ & $11\pm2$ & $14\pm2$ \\
                            & $20$ & $110\pm6$ & $36\pm2$ & $74\pm4$ & $12\pm2$ & $15\pm2$ \\
                            & $30$ & $85\pm4$ & $56\pm3$ & $29\pm1$ & $22\pm3$ & $11\pm2$ \\
                            & $60$ & $58\pm3$ & $50\pm3$ & $7\pm1$ & $37\pm6$ & $16\pm2$ \\
    \hline
    \multirow{5}{*}{BIS-5}  & $0$ & $133\pm7$ & $30\pm2$ & $103\pm5$ & $10\pm1$ & $7\pm1$ \\
                            & $5$ & $128\pm6$ & $35\pm2$ & $93\pm5$ & $12\pm2$ & $6\pm1$ \\
                            & $10$ & $136\pm7$ & $25\pm1$ & $110\pm6$ & $9\pm1$ & $6\pm1$ \\
                            & $20$ & $134\pm7$ & $29\pm1$ & $104\pm5$ & $10\pm2$ & $9\pm1$ \\
                            & $30$ & $127\pm6$ & $36\pm2$ & $91\pm5$ & $12\pm2$ & $10\pm2$ \\
    \hline
    \multirow{8}{*}{BAC-5}  & $0$ & $91\pm5$ & $21\pm1$ & $71\pm4$ & $14\pm2$ & $16\pm2$ \\
                            & $5$ & $97\pm5$ & $0\pm1$ & $97\pm5$ & $9\pm1$ & $13\pm2$ \\
                            & $10$ & $97\pm5$ & $0\pm1$ & $97\pm5$ & $10\pm2$ & $13\pm2$ \\
                            & $20$ & $90\pm5$ & $14\pm1$ & $76\pm4$ & $13\pm2$ & $14\pm2$ \\
                            & $30$ & $88\pm4$ & $10\pm1$ & $78\pm4$ & $14\pm2$ & $15\pm2$ \\
                            & $60$ & $83\pm4$ & $13\pm1$ & $70\pm4$ & $13\pm2$ & $15\pm2$ \\
                            & $120$ & $66\pm3$ & $26\pm1$ & $40\pm2$ & $21\pm3$ & $20\pm3$ \\
                            & $180$ & $71\pm4$ & $16\pm1$ & $54\pm3$ & $20\pm3$ & $20\pm3$ \\
    \hline
  \end{tabular}
\end{table}

\begin{table}[htpb!]
  \caption{Fit parameters of the fuzzy core-shell form factor model obtained for the core-shell microgels (SLS data) after different ultrasonication times $t$.}
  \label{tbl:fit_param_cs}
  \begin{tabular}{ccccccccc}
    \hline
    Microgel & $t$ & $R$ [nm] & $R_{\textrm{c}}$ [nm] & $2\sigma_{\textrm{in}}$ [nm] & $w$ [nm] & $2\sigma_{\textrm{out}}$ [nm] & $\rho_{\textrm{shell}}/\rho_{\textrm{core}}$ & $p$ [\%] \\
    \hline
    \multirow{4}{*}{CS-A}   & $0$ & $350\pm18$ & $38\pm2$ & $209\pm10$ & $58\pm3$ & $45\pm2$ & $0.174\pm0.009$ & $9\pm1$ \\
                            & $5$ & $283\pm14$ & $73\pm4$ & $130\pm6$ & $79\pm4$ & $2\pm1$ & $0.188\pm0.009$ & $10\pm1$ \\
                            & $10$ & $278\pm14$ & $89\pm4$ & $86\pm4$ & $82\pm4$ & $21\pm1$ & $0.20\pm0.01$ & $10\pm1$ \\
                            & $20$ & $267\pm13$ & $107\pm5$ & $44\pm2$ & $101\pm5$ & $15\pm1$ & $0.21\pm0.01$ & $12\pm2$ \\
    \hline
    \multirow{4}{*}{CS-B}   & $0$ & $235\pm12$ & $11\pm1$ & $77\pm4$ & $11\pm1$ & $135\pm7$ & $2.6\pm0.1$ & $13\pm2$ \\
                            & $5$ & $223\pm11$ & $18\pm1$ & $62\pm3$ & $34\pm2$ & $110\pm5$ & $2.0\pm0.1$ & $13\pm2$ \\
                            & $10$ & $202\pm10$ & $16\pm1$ & $71\pm4$ & $40\pm2$ & $74\pm4$ & $2.5\pm0.1$ & $16\pm2$ \\
                            & $20$ & $177\pm9$ & $18\pm1$ & $85\pm4$ & $37\pm2$ & $36\pm2$ & $3.3\pm0.2$ & $18\pm3$ \\
    \hline
  \end{tabular}
\end{table}

\newpage
\section{S6. Additional Atomic Force Microscopy Data}

\begin{figure*}[htpb!]
    \centering
    \includegraphics[width=\linewidth]{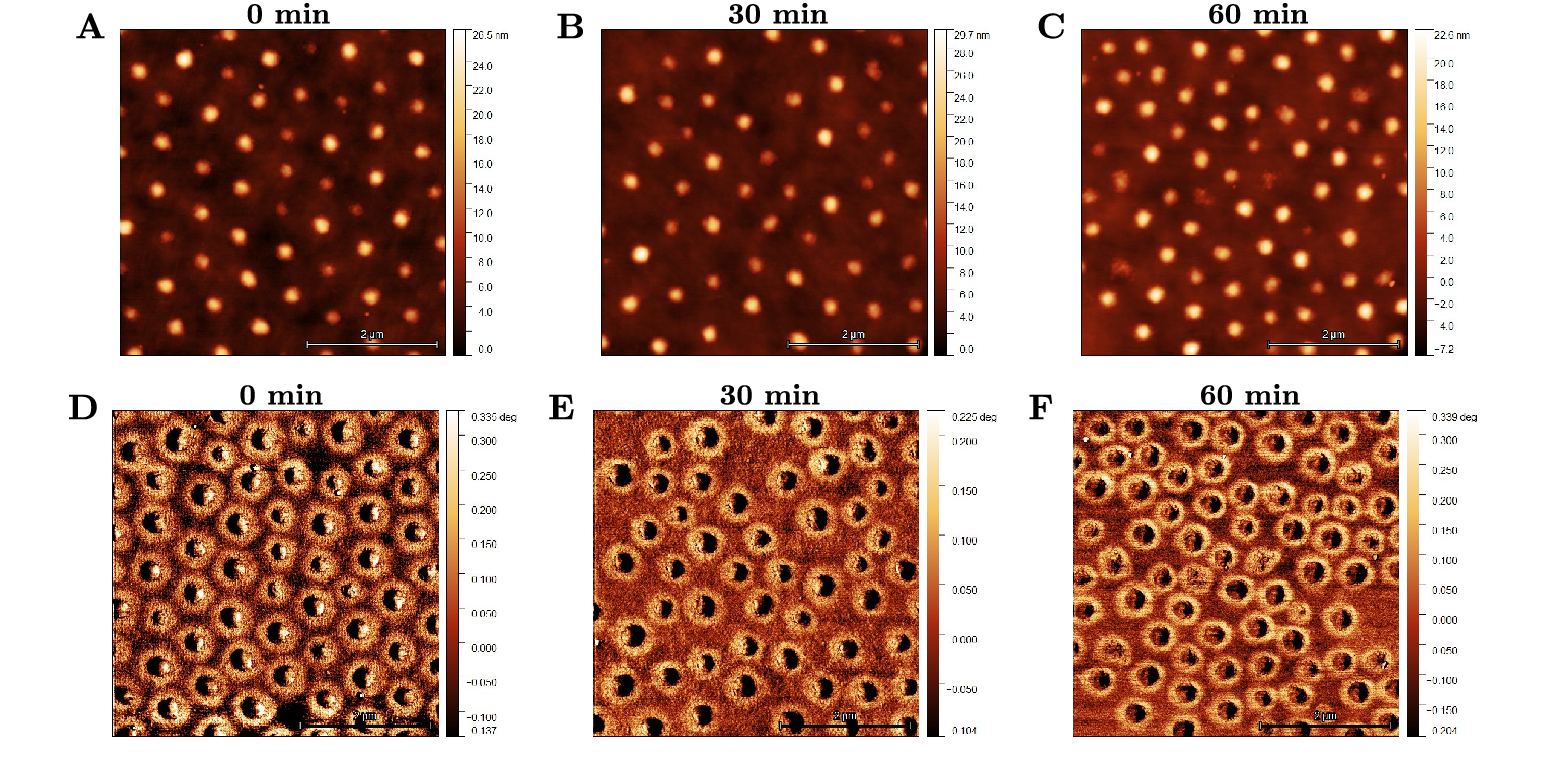}
    \caption{Examples of AFM images of 5\%~BIS microgels after different ultrasonication times (indicated in the figure). Panels A-C correspond to height images, D-F correspond to phase images. Scale bar is 2~$\mu$m.}
    \label{fig:AFM_im_SI}
\end{figure*}

\begin{figure*}[htpb!]
    \centering
    \includegraphics[width=\linewidth]{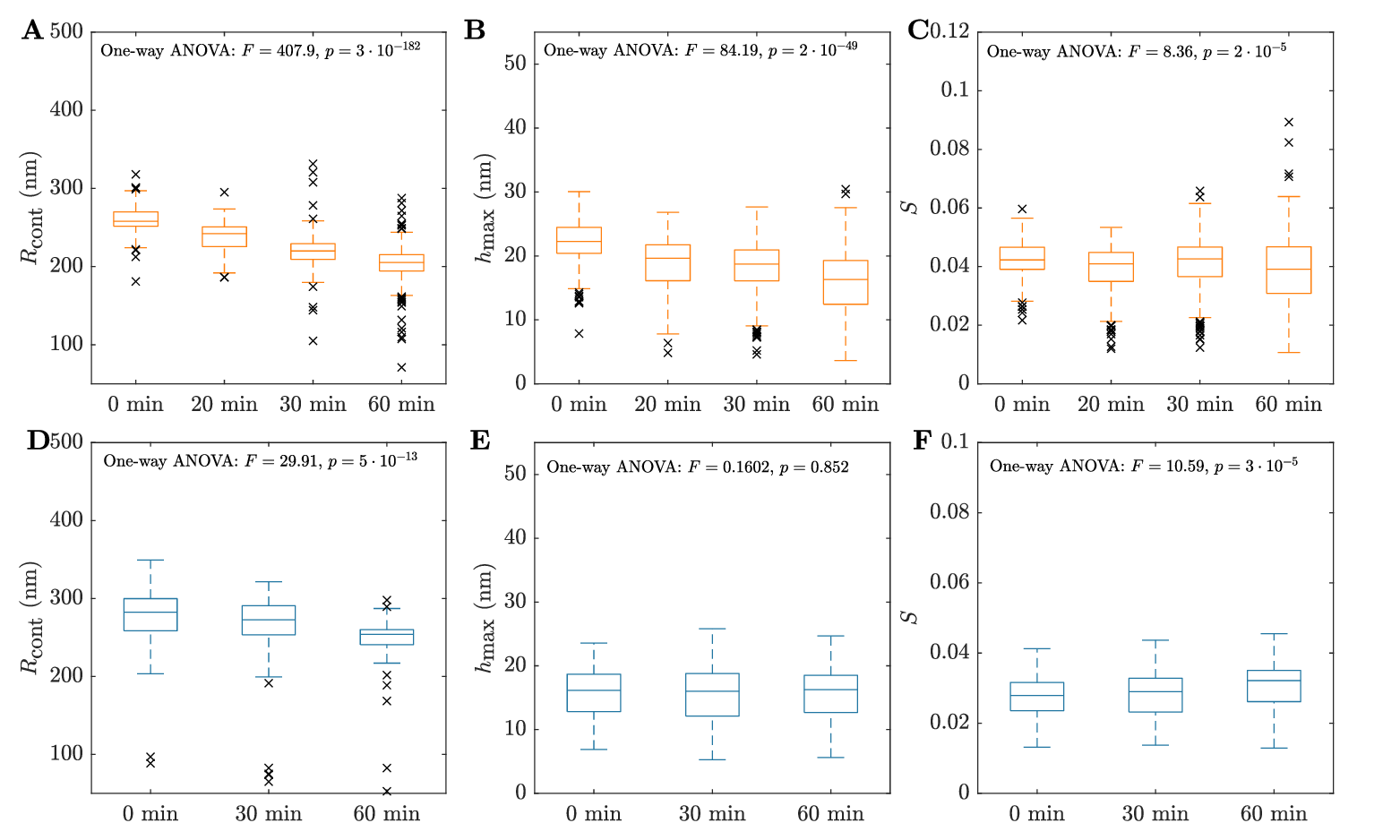}
    \caption{Box plots showing the influence of ultrasonication time $t$ on different parameters obtained by AFM: (\textbf{A}) Contact radii $R_{\textrm{cont}}$ of BAC-1 microgels, (\textbf{B}) Heights $h_{\textrm{max}}$ of BAC-1 microgels, (\textbf{C}) Shape parameter $S$ of BAC-1 microgels, (\textbf{D}) Contact radii $R_{\textrm{cont}}$ of BIS-5 microgels, (\textbf{E}) Heights $h_{\textrm{max}}$ of BIS-5 microgels, (\textbf{F}) Shape parameter $S$ of BIS-5 microgels. Black exes indicates outliers determined as points lying outside of the $\pm2.7\sigma$ interval.}
    \label{fig:AFM_stats}
\end{figure*}

\begin{figure*}[htpb!]
    \centering
    \includegraphics[width=\linewidth]{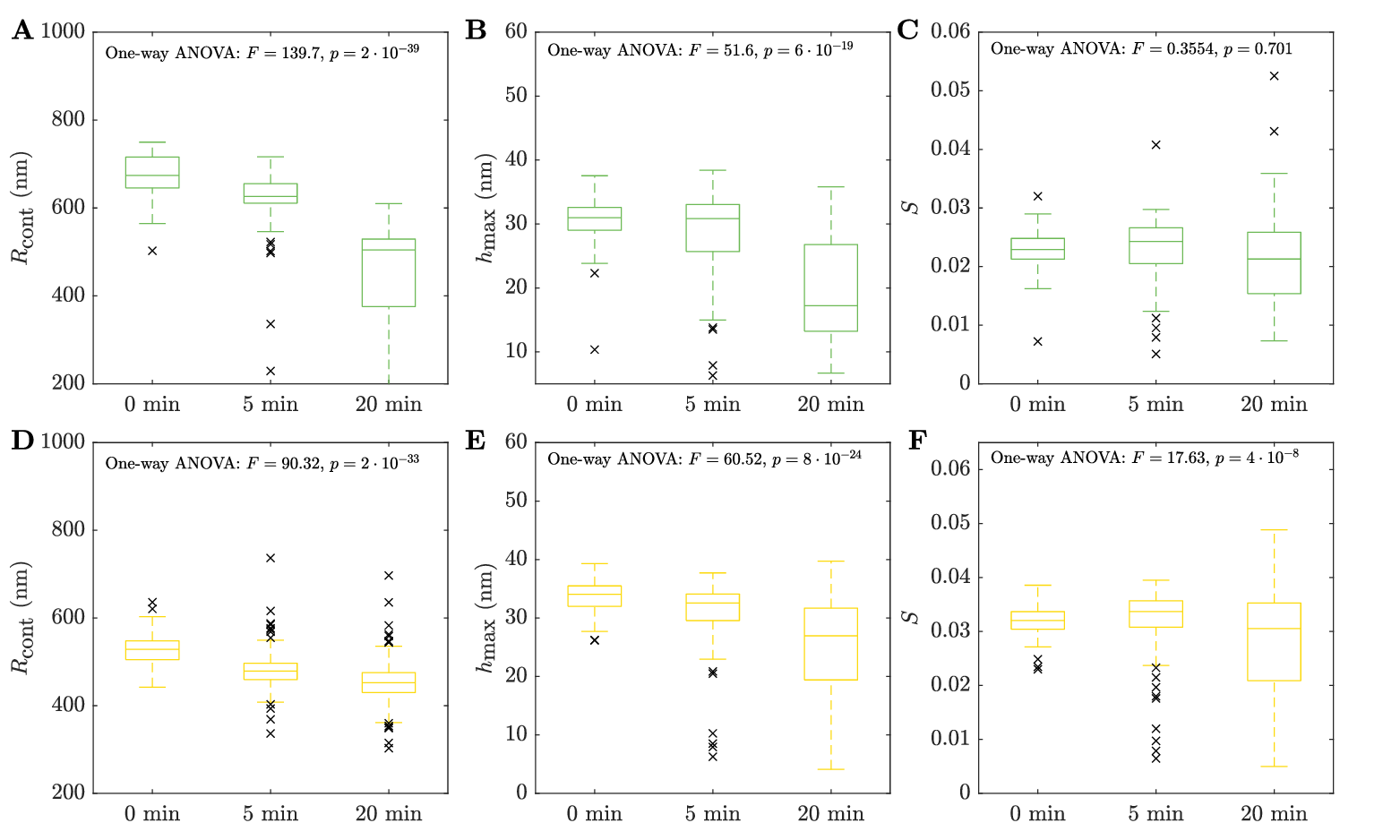}
    \caption{Box plots showing the influence of ultrasonication time $t$ on different parameters obtained by AFM: (\textbf{A}) Contact radii $R_{\textrm{cont}}$ of CS-A microgels, (\textbf{B}) Heights $h_{\textrm{max}}$ of CS-A microgels, (\textbf{C}) Shape parameter $S$ of CS-A microgels, (\textbf{D}) Contact radii $R_{\textrm{cont}}$ of CS-B microgels, (\textbf{E}) Heights $h_{\textrm{max}}$ of CS-B microgels, (\textbf{F}) Shape parameter $S$ of CS-B microgels. Black exes indicates outliers determined as points lying outside of the $\pm2.7\sigma$ interval.}
    \label{fig:AFM_stats_cs}
\end{figure*}

\newpage
\bibliography{refs}